\documentclass[10pt]{article}
\usepackage{graphicx}
\usepackage{dcolumn}
\usepackage{bm}
\usepackage{amssymb,amsmath}
\usepackage[numbers,sort&compress]{natbib}
\begin{document}

\title{Exploring Quantum Control Landscapes:\\Topology, Features, and Optimization Scaling}
\author{Katharine W. Moore and Herschel Rabitz\\Department of Chemistry, Princeton University, Princeton, NJ, 08544}
\date{\today}

\begin{abstract}
Quantum optimal control experiments and simulations have successfully manipulated the dynamics of systems ranging from atoms to biomolecules. Surprisingly, these collective works indicate that the effort (i.e., the number of algorithmic iterations) required to find an optimal control field appears to be essentially invariant to the complexity of the system. The present work explores this matter in a series of systematic optimizations of the state-to-state transition probability on model quantum systems with the number of states $N$ ranging from 5 through 100. The optimizations occur over a landscape defined by the transition probability as a function of the control field. Previous theoretical studies on the topology of quantum control landscapes established that they should be free of sub-optimal traps under reasonable physical conditions. The simulations in this work include nearly 5000 individual optimization test cases, all of which confirm this prediction by fully achieving optimal population transfer of at least 99.9$\%$ upon careful attention to numerical procedures to ensure that the controls are free of constraints. Collectively, the simulation results additionally show invariance of required search effort to system dimension $N$. This behavior is rationalized in terms of the structural features of the underlying control landscape. The very attractive observed scaling with system complexity may be understood by considering the distance traveled on the control landscape during a search and the magnitude of the control landscape slope. Exceptions to this favorable scaling behavior can arise when the initial control field fluence is too large or when the target final state recedes from the initial state as $N$ increases.
\end{abstract}
\maketitle
\section{Introduction}
The control of quantum phenomena with external fields using optimal control theory (OCT) \cite{shi,zhu} and optimal control experiments (OCE) \cite{judson} is currently an active area of research \cite{constantin,raj}. OCT simulations have successfully controlled a variety of objectives, including state preparation \cite{zhu,kral,sola}, molecular isomerization \cite{grossmann,RabO3,abe,artamonov,artamonov2}, dissociation \cite{kaluza,gross, nakagami,abe1}, and orientation/alignment \cite{lapert,salomon,zou}. OCE using ultrafast tailored laser pulses have achieved control over many processes including state preparation \cite{wen,Brixner}, selective molecular dissociation \cite{Baumert, Levis2001, Gerber02}, generation of high order optical harmonics \cite{Bartels2000, Bartels2004, Pfeifer2005}, and energy transfer and isomerization in large biomolecules \cite{herek,Gerber05,Miller06}. Simulation models consider from 2 to $\sim10^2$ or more states, and the atoms/molecules used in OCE often have much larger numbers of accessible states. Remarkably, controlling complex quantum systems appears to be no more difficult than controlling simple ones, both in simulations and experiments, where the level of difficulty is expressed in terms of the number of iterations required to converge on the target objective.

The success of these and other studies suggests that quantum control is generally amenable to ``easy" solution by optimal search. Recently, the {\it quantum control landscape} concept was introduced to help rationalize the observed wide success of quantum control studies \cite{mike1}, where the landscape is defined as the functional relationship between the physical objective (e.g., population transfer probability $P_{i\to f}$) and the external control field $\varepsilon(t)$. Considering a controllable target system under reasonable physical assumptions \cite{ramakrishna}, the topology of the dynamical quantum control landscape can be shown to have no suboptimal local maxima or traps \cite{mike1, demiralp, shen, taksan}. Exceptions to this favorable topology have been found under unusual circumstances, e.g., when constant control fields $\varepsilon(t)$ are employed \cite{schirmer2,tannor2,rebing}. An important objective is to either affirm the attractive theoretical landscape findings or identify the likelihood of encountering landscape traps in the course of typical optimizations under reasonable physical conditions. The extensive prior optimal control literature is supportive of the landscape theory with often high reported yields \cite{shi,zhu,judson,kral,sola,grossmann,RabO3,abe,artamonov,artamonov2,kaluza,gross,nakagami,abe1,lapert,salomon,zou,ambrosek,amstrup,balintkurti,botina,jakubetz,krause,turinici,ohtsuki,ohtsuki2,ohtsuki3,ohtsuki4,phan,ren,schirmer,shah,somloi,wang,wang2,zhu2,zhu3,zhu4}. Such studies, however, cannot rigorously assess the landscape topology due to constraints of various types (e.g., control field fluence) limiting access to the highest yields on the landscape. Additionally, great numerical care is needed when testing the landscape for traps as significant numerical limitations (e.g., insufficient temporal discretization of the control field) can introduce artificial traps. Thus, in the present work we execute a large number of carefully performed numerical simulations to assess the ability to climb the landscape without encountering traps.

This work will consider the control objective of maximizing the probability $P_{i\to f}$ of population transfer from an initial pure state $|i\rangle$ to some target pure state $|f\rangle$ of a closed quantum system undergoing unitary evolution. Although in the laboratory the circumstances will typically include additional factors beyond this idealized situation, the objective of maximizing the population in the product state is often the ultimate goal. The control objective is to identify a suitable field $\varepsilon(t)$ that maximizes $P_{i\to f}$ at some target time $T$, which may be finite or asymptotic with $T\to\infty$. Typically, an optimal field is found using a suitable search algorithm (see, for example, \cite{zhu,tannor}) to traverse the relevant control landscape, which is specified by $P_{i\to f}$ as a functional of the control field, $P_{i\to f}\equiv P_{i\to f}[\varepsilon(t)]$. Both the global topology and local structure of the control landscape may influence the character and duration of the search trajectory from an initial (often random) control field to an optimal solution. 

The search effort required to find an optimal control field is an important issue for determining the feasibility of performing both quantum control simulations and experiments, as computational and experimental resources are inevitably limited. In particular, if the effort rises with system complexity, searching for an optimal control field may become too expensive for complex quantum systems. In this work, the complexity of the system is measured by the Hilbert space dimension $N$, i.e., the number of accessible energy levels of $H_0$. A large body of results from the OCT literature \cite{zhu,judson,RabO3,kaluza,nakagami,lapert,ambrosek,amstrup,balintkurti,botina,jakubetz,krause,turinici,ohtsuki,ohtsuki2,ohtsuki3,ohtsuki4,phan,ren,schirmer,shah,somloi,wang,wang2,zhu2,zhu3,zhu4} performed on systems where $N$ ranges from 2 to $\sim 10^2$ suggest that the search effort required for population transfer does not scale strongly with $N$. Although the required effort will depend on the convergence criteria, the number of reported algorithmic iterations to achieve convergence is observed to be typically no more than $\sim 10^3$, and often $\sim100$ or fewer, regardless of $N$ or the particular search algorithm employed. The invariance of required search effort with respect to $N$ has been numerically demonstrated for the $P_{i\to f}$ objective using so-called $kinematic$ control variables (i.e., the elements of the governing unitary transformation, or equivalent variables) \cite{me}. In the present work, the scaling of the effort with $N$ to find a solution is systematically studied using $dynamic$ control variables (i.e., the control field $\varepsilon(t)$) for simple model quantum systems. Here, effort is defined as the number of algorithmic iterations required to reach a particular threshold value of $P_{i\to f}$; we put aside the effort per iteration to solve the Schr\"{o}dinger equation, which is strongly dependent on $N$. This condition corresponds to the laboratory situation, where the effort of performing an experiment is not necessarily dependent on the complexity of the target molecule. The dynamical control findings throughout the paper will be compared to their kinematic analogs \cite{me}. This comparison is important as similar behavior suggests that the dynamical control behavior has its origins at the simple kinematic level. 

The attractive topology of the quantum control landscape, which will be affirmed in this work, may be expected to contribute to the generally observed favorable lack of scaling of search effort with $N$ \cite{mike1,demiralp}. The attractive global topology, however, does not preclude the possibility that complex local landscape structural features may influence the required search effort, particularly when using a local search procedure such as a gradient algorithm. The high dimensionality of the control landscape (here, the dimensionality is nominally infinite as $\varepsilon(t)$ is a continuous function) renders the direct study of its local structure difficult, but useful information about the local landscape features can be obtained by examining the trajectories taken during a search from an initial to final control. Ultimately, the goal is to understand how the underlying control landscape determines the scaling of the required search effort with $N$.

The remainder of this work is organized as follows. Section \ref{method} formulates the quantum control problem, defines relevant landscape structure metrics, outlines the optimization procedure, and defines the model quantum systems. As a baseline reference to the optimizations, Section \ref{rand} presents the statistical distributions of $P_{i\to f}$ values obtained when random control fields are applied. Section \ref{notraps} shows the important result that no traps were encountered upon optimization of $P_{i\to f}$ in $\sim$5000 test cases. Section \ref{Scale} presents optimization results over varying control targets, Hamiltonians and control fields, with the additional general result that the search effort is invariant to the system complexity characterized by $N$, although the absolute search effort varies widely for different circumstances. In Section \ref{searchstruct}, the effect of landscape features on search effort is explored for the optimal searches performed in Section \ref{Scale} using the metrics defined in Section \ref{method}. Finally, Section \ref{conclusion} presents concluding remarks.

\section{Methods}\label{method}
\subsection{Formulation of the Control Objective}
Consider a quantum system of $N$ levels $|1\rangle,\ldots,|N\rangle$ whose dynamics are driven by the time-dependent Hamiltonian $H(t)=H_0-\mu\varepsilon(t)$, where $H_0$ describes the free dynamics of the system, $\mu$ is the dipole operator, and $\varepsilon(t)$ is the control field. The time-evolution of the quantum system is given by $|\psi(t)\rangle=U(t,0)|\psi(0)\rangle$, where $U(t,0)$ is the unitary evolution matrix covering the dynamics from time $t=0$ to time $t$ and $|\psi(0)\rangle$ is the state of the quantum system at $t=0$. The dynamics of $U$ are governed by the time-dependent Schr\"{o}dinger equation
\begin{equation}
i\hbar\frac{\partial U(t,0)}{\partial t}=H(t)U(t,0),\qquad U(0,0)\equiv \mathbb{I}.\label{sgl}
\end{equation} 

The control objective is to maximize the transition probability $P_{i\to f}$ of population transfer from an initial state $|i\rangle$ to a target state $|f\rangle$ of the system at time $T$,
\begin{equation}
P_{i\to f}(T)\equiv |\langle f|U(T,0)|i\rangle|^2.\label{J}
\end{equation}
The variation of $P_{i\to f}(T)$ with functional changes in the Hamiltonian $H(t)$ is obtained by considering small responses in the propagator $U(t,0)$:
\begin{align}
&i\hbar \frac{\partial}{\partial t}\delta U(t,0)=H(t)\delta U(t,0)+\delta H(t)U(t,0),\qquad \delta U(0,0)=0\label{du}\\
&\delta P_{i\to f}(T)=\langle i|\delta U^{\dag}(T,0)|f\rangle\langle f|U(T,0)|i\rangle+\langle i|U^{\dag}(T,0)|f\rangle\langle f|\delta U(T,0)|i\rangle\label{dp}.
\end{align}
Equation (\ref{du}) can be integrated \cite{taksan} to give
\begin{equation}
\delta U(t,0)=-\frac{i}{\hbar}\int_0^t U(t,t')\delta H(t')U(t',0)dt',\label{intdu}
\end{equation}
and substitution of Eq. (\ref{intdu}) into Eq. (\ref{dp}) yields
\begin{equation}
\delta P_{i\to f}(T)=\frac{2}{\hbar}{\rm Im} \int_0^T\langle i|\delta U^{\dag}(T,0)|f\rangle\langle f|U(T,0)U^{\dag}(t,0)\delta H(t) U(t,0)|i\rangle dt\label{intdp}.
\end{equation}
Within the dipole formulation, $\delta H(t)=-\mu\delta\varepsilon(t)$, which gives the functional derivative $\delta P_{i\to f}/\delta \varepsilon(t)$ from Eq. (6) as
\begin{equation}
\frac{\delta P_{i\to f}}{\delta\varepsilon(t)}=\frac{2}{\hbar}\textup{Im[}\langle i|U^{\dag}(t,0)\mu U(t,0)U^{\dag}(T,0)|f\rangle\langle f|U(T,0)|i\rangle\textup{]}.\label{grad}
\end{equation}

We assume that the system is controllable, such that any arbitrary unitary matrix $U(T,0)$ can be generated by a suitably chosen field $\varepsilon(t)$ at a sufficiently large final time $T$. This condition is equivalent to the requirement that the Lie algebra generated from $H_0$ and $\mu$ forms a complete set of operators \cite{ramakrishna} and $T$ is large enough to avoid hindering the dynamics. In general, we may assume controllability of an arbitrary quantum system, as uncontrollable quantum systems have been shown to constitute a null set in the space of Hamiltonians \cite{altafini}. Upon satisfaction of the controllability requirement, analysis of the global control landscape topology of Eq. (\ref{J}) with kinematic variables \cite{mike1} reveals that the landscape has no false extrema; the only critical points occur at perfect control, $P_{i\to f}=1$, and no control, $P_{i\to f}=0$. Upon satisfaction of the Jacobian $\delta U(T,0)/\delta\varepsilon(t)$ being full-rank, the dynamical landscape also has no traps \cite{raj,taksan} and the desired landscape value $P_{i\to f}=1$ corresponds to a submanifold of optimal fields, which makes the control solutions robust to fluctuations in $\varepsilon(t)$ \cite{demiralp,shen}. The latter property is particularly important for laboratory quantum control, as it allows for maintaining good yields despite laboratory noise. In practice, the rank of $\delta U(T,0)/\delta\varepsilon(t)$ may be reduced to some degree with no impact on the controlled dynamics, as there can still be many readily traversed pathways from $|i\rangle$ to $|f\rangle$. However, traps may arise for so-called singular control fields where the above Jacobian is significantly rank-deficient. Such situations have been known to occur when $\varepsilon(t)=$ constant is employed \cite{schirmer2,tannor2,rebing}, but this situation is generally not physically relevant in the laboratory. Thus, one goal of the simulations in this work is to establish whether traps may be encountered in optimizations starting from physically reasonable control fields.

\subsection{Measuring Landscape Structure}\label{struct}
The {\it global landscape topology} summarized above provides important information about the feasibility of achieving optimal control. The claimed lack of traps means that a control producing a perfect yield can be found starting from any initial search point on the landscape (i.e., a point on the landscape corresponds to a particular field and its associated transition probability) using a suitable hill-climbing algorithm. The validity of this topology in OCT simulations will be assessed in this work.

The presence of a favorable landscape topology does not preclude the presence of increasingly complex landscape features as $N$ rises, which could cause an increase in the search effort to find a control that gives perfect yield. Thus, an understanding of {\it local landscape features} (i.e., non-critical point structures) is necessary in order to explain and predict the scaling of search effort with system complexity. In this work, the local features of the control landscape are codified by specific metrics recorded along the search trajectory followed from the initial to optimal control field. On a given search trajectory, we may parametrize the field $\varepsilon(t)$ by an index $s\geq0$ to track the progress to the top of the landscape. The field starts out at $s$=0 with $\varepsilon(0,t)$ and progresses in steps $s\to s+ds$ (i.e., $\varepsilon(s,t)\to\varepsilon(s+ds,t)$) until the trajectory ends at an optimal control, $\varepsilon_{opt}=\varepsilon(s_{M},t)$ at $s=s_{M}$.

For the purpose of describing the local landscape features, we define (i) a distance metric between two fields $\varepsilon(s,t)$ and $\varepsilon(s',t)$ ($t\in[0,T]$) based on $||\varepsilon(s,t)-\varepsilon(s',t)||$, where $||\cdot||$ implies an integration over time, and (ii) structure metrics based on a Taylor expansion of $P_{i\to f}$ around a field $\varepsilon(s,t)$ at points on the landscape. Analogous metrics of local landscape features were defined in \cite{me} using kinematic control variables (i.e., without reference to the dynamics of any particular Hamiltonian) and were found to correlate with the observed scaling of the search effort with $N$. From this experience, these metrics are used here to provide information about how the features of the landscape determine the required search effort using dynamic variables. 

The complexity, or gnarled character, of a search trajectory in control space must take into account both the Euclidian distance between the initial and final control fields and the actual path length followed from the initial to final control over the course of a search. A metric defining this complexity may be characterized by the ratio of the trajectory path length $||\Delta_P\varepsilon(t)||$ to the Euclidian distance between the initial and final control fields $||\Delta_E\varepsilon(t)||$,
\begin{align}
R_{\varepsilon}&=\frac{||\Delta_P\varepsilon(t)||}{||\Delta_E\varepsilon(t)||}=\frac{\int_0^{s_{M}}ds\left(\int_0^T dt \left[\frac{d\varepsilon(s,t)}{ds}\right]^2\right)^{1/2}}{\left(\int_0^T dt \left[\varepsilon(s_{M},t)-\varepsilon(0,t)\right]^2\right)^{1/2}}\geq1\label{ratio}
\end{align}
The closer $R_{\varepsilon}$ is to unity, then the more direct the path, i.e., the closer the path is to a straight line in the space of controls being searched over. Following a direct path from the initial to optimal control field should result in efficient searching, especially by simple local algorithms, because the search trajectory could avoid taking detours along the way to finding an optimal control field. This prediction will be assessed in the simulations.

The local structure metrics of the landscape provide information about what the search algorithm ``sees" at a particular point on the landscape and may be expressed through a Taylor expansion of the cost functional $P_{i\to f}$,  
\begin{equation}
P_{i\to f}[\varepsilon(s,t) + \delta \varepsilon(s,t)] = P_{i\to f}[\varepsilon(s,t)] + \int_0^T \nabla P_{i\to f}(s,t)\delta\varepsilon(s,t)dt + \frac{1}{2}\int_0^T \int_0^T \mathcal{H}(t,t')\delta\varepsilon(s,t)\delta\varepsilon(s,t')dtdt'+\cdots\label{taylor},
\end{equation}
where $\nabla P_{i\to f}(s,t)=\delta P_{i\to f}/\delta\varepsilon(s,t)$ is the gradient vector. The structure metrics will be extracted from the kernels of the integrals in Eq. \ref{taylor}. Each metric will be labelled by $m$ to indicate its evaluation at the point $s_m$ on the landscape. The first-order term in Eq. (\ref{taylor}) specifies the slope metric $\mathcal{S}_m$,
\begin{equation}
\mathcal{S}_{m}=\Bigr\rvert\Bigr\rvert\nabla P_{i\to f}(s_m,t)\Bigr\rvert\Bigr\rvert= \left(\int_0^T dt\left(\frac{\delta P_{i\to f}}{\delta\varepsilon(s_m,t)}\right)^2\right)^{1/2}.\label{steep}
\end{equation}
The slope metric is equivalent to the magnitude of the gradient on the landscape at the point $s_m$. Intuitively, a greater value of $\mathcal{S}_m$ should result in a locally faster ascent due to a more rapid improvement of the yield when taking a step in the direction of the gradient. Thus, it is expected that the slope metric may be correlated to the observed search effort. 

Additional information about local landscape features can be gained by examining the second-order term of the Taylor expansion in Eq. (\ref{taylor}), or the Hessian matrix, whose elements labelled by $t$ and $t'$ are \cite{demiralp}
\begin{align}
{\cal H}(t,t')=\frac{\delta^2 P_{i\to f}}{\delta\varepsilon(t)\delta\varepsilon(t')}=&2{\rm Re}[\langle i|U(0,T)|f\rangle\langle f|U(T,t)\mu U(t,t')\mu U(t',0)|i\rangle\nonumber\\
&-\langle i|U(0,t)\mu U(t,T)|f\rangle\langle f|U(T,t')\mu U(t',0)|i\rangle],\quad t\geq t'.
\end{align}
The Hessian matrix is symmetric, i.e., ${\cal H}(t,t')\equiv{\cal H}(t',t)$. Two simple metrics based on the Hessian matrix can provide insight into the landscape structure, particularly at the bottom and top of the landscape. The first metric is the Hessian trace,
\begin{equation}
{\rm Tr}{\cal H}=\int_0^T{\cal H}(t,t)dt,\label{trh}
\end{equation}
and the second metric is the curvature of the landscape at a point $m$,
\begin{equation}
{\cal C}_m=\left(\frac{1}{\Bigr\rvert\Bigr\rvert\nabla P_{i\to f}(s_m,t)\Bigr\rvert\Bigr\rvert}\right)^2\int_0^Tdt\int_0^Tdt'\nabla P_{i\to f}(s_m,t)^{\dag}{\cal H}(t,t')\nabla P_{i\to f}(s_m,t'),\label{curve}
\end{equation}
which may be calculated anywhere including near, but not at, the bottom or top of the landscape where $\nabla P_{i\to f}(s_m,t)=0$. The curvature defined by Eq. (\ref{curve}) is the Hessian projected along the normalized local gradient direction. Intuitively, a larger (positive) value of the Hessian trace and curvature near the bottom of the landscape should induce fast climbing \cite{vinny2}. Similarly, a large (negative) value of the curvature ${\cal C}_m$ and Hessian trace Tr$\cal H$ near the top should also accelerate the approach to the optimum.

\subsection{Optimization procedure}\label{alg}
Many different search algorithms may be used to find an optimal field $\varepsilon(s_{M},t)$ maximizing $P_{i\to f}$. One important goal of this work is to assess whether traps are encountered upon climbing the landscape; the existence of traps could preclude identification of an optimal control field producing $P_{i\to f}\sim1.0$. This landscape assessment objective specifically calls for a local (i.e., myopic) search method, which will stop climbing at a sub-optimal value of $P_{i\to f}$ if a trap is encountered. Global search algorithms (e.g., genetic algorithms) may step over traps, making them inappropriate for assessing topology. Additionally, the particular choice of search algorithm may significantly influence the absolute effort required to find an optimal field; this was found to be the case for optimizing $P_{i\to f}$ using kinematic controls \cite{me}, where gradient, genetic, simplex, and coordinate search algorithms were compared. Despite the wide variation in absolute search effort with the choice of algorithm, the $scaling$ of the search effort with respect to system complexity exhibited the same qualitative trends for all algorithms examined. Similarly, in OCT studies from the literature, gradient-based algorithms typically converge in $\sim 100$ iterations \cite{zhu,ambrosek,RabO3,balintkurti,botina,jakubetz,krause,lapert,turinici,ohtsuki,ohtsuki4,ren,shah,somloi,wang,wang2,zhu2,zhu3,zhu4}, while non-gradient simplex and evolutionary searches typically require several hundred iterations \cite{judson,amstrup,kaluza}. Importantly, these numbers do not appear strongly dependent on $N$. Considering all of the factors above, a gradient algorithm is employed exclusively in this work in order to (a) test the likelihood of encountering traps, and (b) seek consistency in exploring optimization effort.

As the control field $\varepsilon(s,t)$ depends on the variable $s$ labeling the progression of the optimization, the landscape value $P_{i\to f}(s)\equiv P_{i\to f}[\varepsilon(s,t)]$ depends on $s$ through its functional dependence on $\varepsilon(s,t),\medspace 0\leq t\leq T$. Thus, the change in the landscape value $P_{i\to f}$ corresponding to a differential change $ds$ is given by $dP_{i\to f}\equiv \left(\frac{\partial P_{i\to f}}{\partial s}\right)ds$, where
\begin{equation}
\frac{dP_{i\to f}}{ds}\equiv \int_0^T dt \frac{\delta P_{i\to f}}{\delta\varepsilon(s,t)}\frac{\partial \varepsilon(s,t)}{\partial s}\label{dJds}
\end{equation} 
As the objective is to maximize $P_{i\to f}$, we have the demand that $\frac{dP_{i\to f}}{ds}>0$, so $\varepsilon(s,t)$ satisfies the differential equation 
\begin{equation}
\frac{\partial\varepsilon(s,t)}{\partial s}=\frac{\delta P_{i\to f}}{\delta\varepsilon(s,t)},\label{ode}
\end{equation}
where the gradient on the right-hand side is given by Eq. (\ref{grad}). Carefully solving Eq. (\ref{ode}) coupled to the Schr\"odinger equation (\ref{sgl}) is essential for obtaining reliable landscape climbing results, especially for assessing the presence of traps. The present search algorithm, incorporated into MATLAB \cite{matlab}, solves Eq. (\ref{ode}) using a fourth order Runge-Kutta integrator with a variable step size to determine the control field at the next iteration. Of additional special interest here is the required search effort, or the number of algorithmic iterations $M$ required to reach the desired $P_{i\to f}$ value, when starting from an initial random control field.

\subsection{Design of quantum systems for simulations}\label{ham}
The goals of the simulations are to (a) assess whether traps are encountered in carefully performed optimizations and (b) explore general trends in the scaling of search effort to find optimal controls in relation to system complexity. For a proper assessment of goal (a), as well as in the simulations for (b), no fluence or other direct constraints are placed on the controls, aside from a fine time discretization of the field. Since an infinite variety of structures for $H_0$ and $\mu$ can arise, a thorough sampling of all physically relevant structures is infeasible. Nevertheless, a modest number of variations in $H_0$, $\mu$, and choice of $|i\rangle$ and $|f\rangle$ can capture broad classes of physical phenomena. Increasing $N$ while holding the $|i\rangle\to|f\rangle$ target transition fixed corresponds to exciting the same transition in homologous molecules of increasing size. The circumstance of fixing $N$ and the target transition while varying the dipole matrix structure corresponds to controlling homologous molecules of similar size with different transition couplings. Choosing the target transition as $|1\rangle\to|N\rangle$ and increasing $N$ corresponds to exciting larger molecules to an ever receding highest quantum level. In practice, the target $|i\rangle\to|f\rangle$ transition, dipole matrix structure, and $N$ will likely vary simultaneously in the laboratory. The results here should both provide diverse test scenarios for the presence of landscape traps as well as capture the qualitative search effort scaling trends. Comparisons to the corresponding laboratory situations will be made at relevant points throughout the work.

For all of the simulations in this work, we consider an $N$-level quantum system whose Hamiltonian is expressed in arbitrary dimensionless units. Two general choices of nondegenerate, diagonal $H_0$ are employed, corresponding qualitatively to a rigid rotor or an anharmonic oscillator. The energy levels of the rigid rotor are given by 
\begin{equation}
H_0=\sum_{j=0}^{N-1} \gamma \thinspace j\left(j+1\right)|j\rangle\langle j|,\label{ho}
\end{equation}
where $\gamma$ is a constant. In the results presented here, $\gamma=0.25$, but varying $\gamma$ was found to have no significant effect on the scaling of search effort with $N$ or on the local landscape structure. The energy levels of the anharmonic oscillator are
\begin{equation}
H_0=\sum_{j=0}^{N-1}\left[\omega\left(j+\frac{1}{2}\right)-\frac{\omega^2}{\cal D}\left(j+\frac{1}{2}\right)^2\right]|j\rangle\langle j|\label{osc},
\end{equation}
where $\omega=5$ and ${\cal D}=1200$ for all results presented here. Variation of $\omega$ and $\cal D$ were found not to affect the search effort scaling, provided that they were chosen to allow for significantly more bound states than the value of $N$ employed in the simulations. The above choices of $\omega$ and $\cal D$ provide 120 bound states. The $H_0$ structures given in Eqs. (\ref{ho}) and (\ref{osc}) will be referred to respectively as the rotor and oscillator $H_0$ structures later.

Two physically relevant dipole real matrix structures will be considered. For many physical systems the coupling between states generally decreases as the difference between the quantum numbers of the states increases, and the present choices of $\mu$ take this property into account. We first choose $\mu$ to have the simple structure
\begin{equation}
\mu=
\begin{pmatrix}0& 1&D& D^2& \ldots& D^{N-2}\\
1& 0& 1& D& \ldots& D^{N-3}\\
D& 1& 0& 1& \ldots& D^{N-4}\\
D^2& D&1& 0&\ldots& D^{N-5}\\
\vdots& \vdots& \vdots& \vdots& \ddots& \vdots\\
D^{N-2}&D^{N-3}& D^{N-4}& D^{N-5}&\ldots& 0
\end{pmatrix}\label{mu}
\end{equation}
where $D\in[0,1]$ is the drop-off rate and all elements of $\mu$ have a random phase of $\pm 1$ with the restriction that $\mu$ remains symmetric. We further specify that $\mu_{if}=0$, thereby eliminating a direct transition from the initial state $|i\rangle$ to desired target state $|f\rangle$.

In order to generalize the structure of $\mu$ from that shown in Eq. (\ref{mu}), we alternatively chose $\mu$ to have the form.
\begin{equation}
\mu=
\begin{pmatrix}0& \alpha_1&\alpha_2& \alpha_3& \ldots& \alpha_{N-1}\\
\alpha_1& 0& \alpha_1& \alpha_2& \alpha_3& \vdots\\
\alpha_2& \alpha_1& 0& \alpha_1& \alpha_2& \vdots\\
\alpha_3& \alpha_2&\alpha_1& 0&\alpha_1& \vdots\\
\vdots& \vdots& \vdots& \vdots& \ddots& \vdots\\
\alpha_{N-1}&\ldots& \alpha_3&\alpha_2&\alpha_1& 0
\end{pmatrix}\label{mu2}
\end{equation}
The successive superdiagonal elements $\alpha_i$, $i=1\ldots N-1$, are each chosen from particular uniform random distributions such that $\alpha_1\in[0.8,1]$, $\alpha_2\in[0.7,0.9]$, $\alpha_3\in[0.6,0.8]$, $\ldots$ $\alpha_{i\geq10}\in[0,0.1]$. While preserving symmetry, all nonzero elements have a random phase of $\pm1$, and $\mu_{if}=0$. The choice of the dipole matrices in Eqs. (\ref{mu}) and (\ref{mu2}), respectively, will be referred to the $D$ and $\alpha$ structures later. The freedom inherent in randomly drawing the coupling matrices provides a broad family of systems to assess the landscape topology, structural features, and search effort scaling behavior.

In many OCT studies, the initial control field is chosen based on knowledge of the physical system. For example, the component spectral frequencies are often picked to be resonant with certain transitions in $H_0$, or a spectral bandwidth is chosen that encompasses the desired transitions. In this work, the initial electric field $\varepsilon(0,t)$ is discretized on a time interval $t\in[0,28]$ into $2048$ time-points. The choice of $T$=28 and 2048 discretized time-points was found to be sufficient to resolve the fastest modulation in the field $\varepsilon(s,t)$ and the fastest modulation in the wavefunction $|\psi(t)\rangle$ for all systems of $N<30$. For simulations involving the $|1\rangle\to|N\rangle$ transition for $N\geq 30$, 4096 time points were used to ensure sufficient resolution. 

The initial field at $s=0$ is chosen as  
\begin{equation}
\varepsilon(0,t)=F\textup{exp}\left[-\beta\left(t-\frac{T}{2}\right)^2\right]\sum_{k=1}^{K}\textup{sin}\left(\omega_kt+\phi_k\right),\thickspace t\in[0,\thinspace T]
\end{equation}
where $\beta$ is an envelope parameter (in all simulations, $\beta$=0.05), $K$ is the number of frequency components, $\phi_k$ is a random phase on $[0,2\pi]$, and $F$ is the square root of the field fluence. Prior to multiplication by $F$, the field is normalized to have unit fluence. The frequencies $\{\omega_k\}$ are chosen randomly on a pre-defined bandwidth with maximal frequency $\Omega$. In most simulations, $\Omega$ corresponds to the frequency of the $|1\rangle\to|f\rangle$ transition in $H_0$, but in Section \ref{field}, other choices of $\Omega$ are employed. Following selection of the initial frequencies $\{\omega_k\}$ and the field fluence $F$, the electric field is allowed to vary freely over the optimization in terms of each of its time-points $\varepsilon(s,t_j),\medspace j=1,2,\ldots, 2048$ (or $\varepsilon(s,t_j),\medspace j=1,2,\ldots, 4096$ for some cases where $N\geq 30$) as control variables starting at $s=0$ and iteratively moving ahead as $s\to s+\Delta s$. 

\section{Statistical Distribution of $P_{i\to f}$ Yields}\label{rand}
It is instructive to examine the statistical distribution of $P_{i\to f}$ values upon making random choices for the initial control field $\varepsilon(0,t)$ because many OCE searches for effective controls start with a random trial choice. Of particular interest is whether the optimization searches, on average, start at more or less favorable landscape values as $N$ increases. 

A detailed mathematical analysis of the $P_{i\to f}$ objective with kinematic controls shows that the statistics satisfy a $\beta$-distribution \cite{mike3}. As $N$ increases, this distribution becomes skewed towards smaller $P_{i\to f}$ values. This qualitative behavior has also been observed for initial choices of random control fields $\varepsilon(0,t)$ \cite{shen} for the target transition $|1\rangle\to|N\rangle$. Therefore, simply considering the statistical distribution for random trials suggests that increasing search difficulty may be encountered as $N$ grows. In order to systematically test the validity of this conjecture under different initial conditions, we chose (a) target transitions $|1\rangle\to|5\rangle$, $|1\rangle\to|10\rangle$, and $|1\rangle\to|N\rangle$, (b) control field fluence $F$=10, 1, 0.1 and (c) dipole matrices of structure $D$ in Eq. (\ref{mu}) with $D$=0.5, 0.2, for $N$ ranging from 5 to 40. The statistics were obtained for 10$^4$ different randomly generated control fields for each set of parameters $|1\rangle\to|f\rangle$, $D$, and $F$. All control fields had $K$=20 frequencies randomly distributed on the bandwidth with maximal frequency $\Omega$=$\omega_{f}$, where $\omega_f$ denotes the frequency corresponding to the $|1\rangle\to|f\rangle$ transition. Results using the rotor Hamiltonian given in Eq. (\ref{ho}) are shown here; choice of the oscillator Hamiltonian in Eq. (\ref{osc}) produced qualitatively similar results.

\begin{figure}
\includegraphics[width=8.5cm]{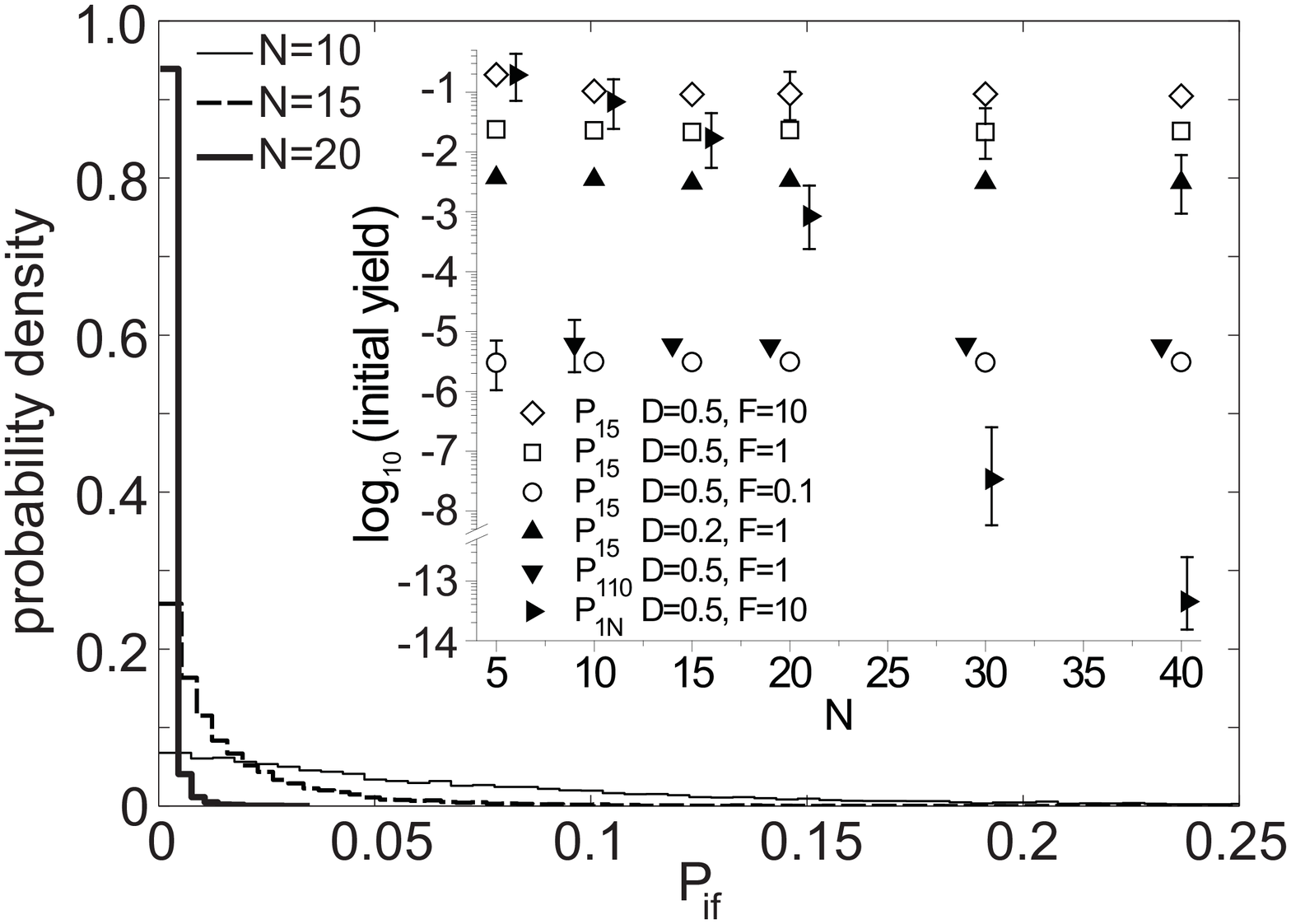}
\caption{\label{stats}}
\end{figure}

Figure \ref{stats} presents the distribution functions for the $|1\rangle\to|N\rangle$ transition with fields of $F$=10 for $N$=10, 15, and 20, revealing a shift towards reduced values of $P_{i\to f}$ as $N$ rises. The inset of Figure \ref{stats} shows the mean of the statistical distribution versus $N$ for the cases of different targets, field strengths, and dipole matrix drop off rates as labeled in the legend, where $D$ and $F$ are denoted for each $P_{i\to f}$ target. For any fixed target transition (e.g., $|1\rangle\to|5\rangle$), the mean of each distribution is independent of $N$, and the distributions for these cases are indistinguishable as $N$ is varied (not shown). For the $|1\rangle\to|N\rangle$ transition, the mean $P_{i\to f}$ value decreases rapidly with rising $N$ (note the log scale), in accordance with \cite{MikeRebing}, indicating that it becomes increasingly difficult to find a decent initial yield as $N$ rises for the receding target $|N\rangle$. The average initial yield for systems with a fixed target transition, however, should not change dramatically as system dimension rises. Instead, the fluence of the initial control field and dipole coupling strength appear to determine the initial yield, with the trends following intuitive insights. As expected, stronger fields result in a greater yield than weaker fields; however, at very strong fields (not shown), this trend can reverse due to amplitude spreading over all the states. Similarly, lower yields are obtained for systems with weaker coupling indicated by smaller $D$ values. 

\section{Testing for the Presence of Traps on the Landscape}\label{notraps}
Of primary importance for the utility of quantum OCT and OCE is the question of whether all searches starting from a random initial field $\varepsilon(0,t)$ can even find an optimal field achieving $P_{i\to f}\sim$1 without getting trapped at a suboptimal $P_{i\to f}$ value. Under reasonable assumptions, the topology of the control landscape has been theoretically shown to contain no suboptimal extrema when the system is controllable, no constraints are placed on the controls, and the Jacobian $\delta U(T,0)/\delta\varepsilon(t)$ is full-rank \cite{raj,mike1,demiralp,taksan}. Affirming this attractive topological prediction is very important, as special instances of traps can be found \cite{schirmer2,tannor2,rebing} under unusual conditions. For the $P_{i\to f}$ objective, the OCT literature regularly reports excellent results \cite{shi,zhu,judson,kral,sola,grossmann,RabO3,abe,artamonov,artamonov2,kaluza,gross,nakagami,abe1,lapert,salomon,zou,ambrosek,amstrup,balintkurti,botina,jakubetz,krause,turinici,ohtsuki,ohtsuki2,ohtsuki3,ohtsuki4,phan,ren,schirmer,shah,somloi,wang,wang2,zhu2,zhu3,zhu4}, with maximum yields of $P_{i\to f}\simeq0.9$ or greater. These results are not definitive for fully testing the landscape theory, as fluence or other field constraints are typically present, and special computational care may be required to eliminate artificial traps due to numerical aberrations. The present calculations paid due attention to all such details to provide a large-scale test of the landscape topology predictions for $P_{i\to f}$. As pointed out in Section \ref{alg}, a gradient-based algorithm was used because a local search will stop if a trap is encountered. It is important to execute the gradient algorithm in a stable fashion for this purpose, so a fourth-order Runge-Kutta procedure was employed.

{\it This work provides broad systematic evidence that optimization searches can achieve a high yield of $P_{i\to f}\geq$0.999 without encountering suboptimal extrema.} A total of $\sim$5000 individual optimal searches were performed with a wide variety of control parameters chosen (c.f., Section \ref{ham}) for $N$ ranging from 5 to 100. In order to ensure that no false traps resulted from choices of simulation parameters, the control field was allowed to have as much fluence as necessary and the final time $T$ was chosen to be sufficiently large so as not to impose a constraint. The importance of paying proper attention to all numerical details was evident for some of the difficult cases with the target transition of $|1\rangle\to|N\rangle$ (see Section \ref{1n} for further details) when $N$=30 and 40 with the $D$=0.5 dipole, the rotor $H_0$, and employing 2048 time-points to discretize the control field. Out of the $\sim$5000 tests, 12 of the latter category were ``trapped'' at yields of $0.997-0.998$. However, upon interpolation of the trapped control fields on 4096 time-points and continued ascent with the gradient algorithm, the demanded criterion of $P_{i\to f}\geq$0.999 was achieved in these cases. Similar results were observed for optimization of the control objective of generating a target unitary transformation $U(T,0)$ with a control field $\varepsilon(t)$ to match some target unitary matrix $W$. This objective may be measured by considering the fidelity function $J=||W-U(T,0)||^2$. In the latter study, 20,000 tests were performed on quantum systems with 2-16 energy levels; upon choice of a sufficiently fine time-mesh and large $T$, each optimization converged to a fidelity value of $J\leq10^{-6}$ \cite{mew}.

Collectively, these results indicate that the likelihood of finding traps on quantum control landscapes is vanishingly small when starting with reasonable control fields, allowing access to sufficiently flexible controls, and paying attention to numerical details. This result suggests that the traps in \cite{schirmer2,tannor2,rebing} are at most an extremely rare occurrence on the landscape, and possibly a null set. Another consideration is that many practical OCT and OCE studies may be considered as quite successful upon even reaching moderate yields when operating with various constraints. Importantly, the landscape principles affirmed by the tests here imply that under such conditions the enhancement of control resources can open up even higher yields.

\section{Search Effort and System Complexity}\label{Scale}
The scaling of the required search effort with system complexity can determine the feasibility of performing quantum control on polyatomic molecules or similarly complex systems. Intuitively, the expectation is that finding a suitable control field would become more difficult as the size of the system increases, because additional control pathways involving a larger number of quantum states become accessible. The collective OCT literature, however, suggests that the required search effort to find an optimal control is generally on the order of $\sim10^2$ iterations, \cite{zhu,judson,RabO3,kaluza,nakagami,lapert,ambrosek,amstrup,balintkurti,botina,jakubetz,krause,turinici,ohtsuki,ohtsuki2,ohtsuki3,ohtsuki4,phan,ren,schirmer,shah,somloi,wang,wang2,zhu2,zhu3,zhu4}, and systematic optimization of $P_{i\to f}$ using kinematic control variables indicates that the search effort scales at most very slowly with $N$ \cite{me}. Successful OCE studies ranging from control of atoms \cite{wen,Bartels2000} to complex protein molecules \cite{herek,Miller06} further suggest a practical level of invariance of search effort to system complexity. Based on these collective findings, we performed optimization of $P_{i\to f}$ on a broad sampling of systems ranging from $N$=5 to $N$=100 in order to determine whether scaling invariance to $N$ can be demonstrated systematically using dynamical control variables. The effects of changing the dipole coupling strength, the control field parameters, and the $|i\rangle\to|f\rangle$ target transition on the search effort and its scaling with $N$ are examined here. 

\subsection{Varying Dipole Coupling Strength}\label{dip}
Optimizations were performed for systems with $N$ ranging from 5 to 40 as well as $N$=100 for the target transitions $|1\rangle\to|5\rangle$ and $|1\rangle\to|10\rangle$. Dipole structures of $D$=1.0, 0.5, and 0.2 as well as the $\alpha$ structure were examined, with $H_0$ given by Eq. (\ref{ho}) or Eq. (\ref{osc}). For all simulations, the initial control fields of $F$=1 had $K$=20 frequencies randomly chosen on a bandwidth with maximal frequency $\Omega$ corresponding to the $|1\rangle\to|f\rangle$ transition in $H_0$. Optimal searches beginning from 20 such initial fields were performed for each choice of $N$ and dipole structure, with the exception of $N$=100, where 10 optimal searches were performed. In order to normalize the reported search effort with respect to the initial $P_{i\to f}$ yields obtained, the counting of iterations was begun at $P_{i\to f}\geq$0.001, regardless of the initial yield, and random fields producing $P_{i\to f}\geq$0.01 were discarded.

\begin{figure}
\includegraphics[width=8.5cm]{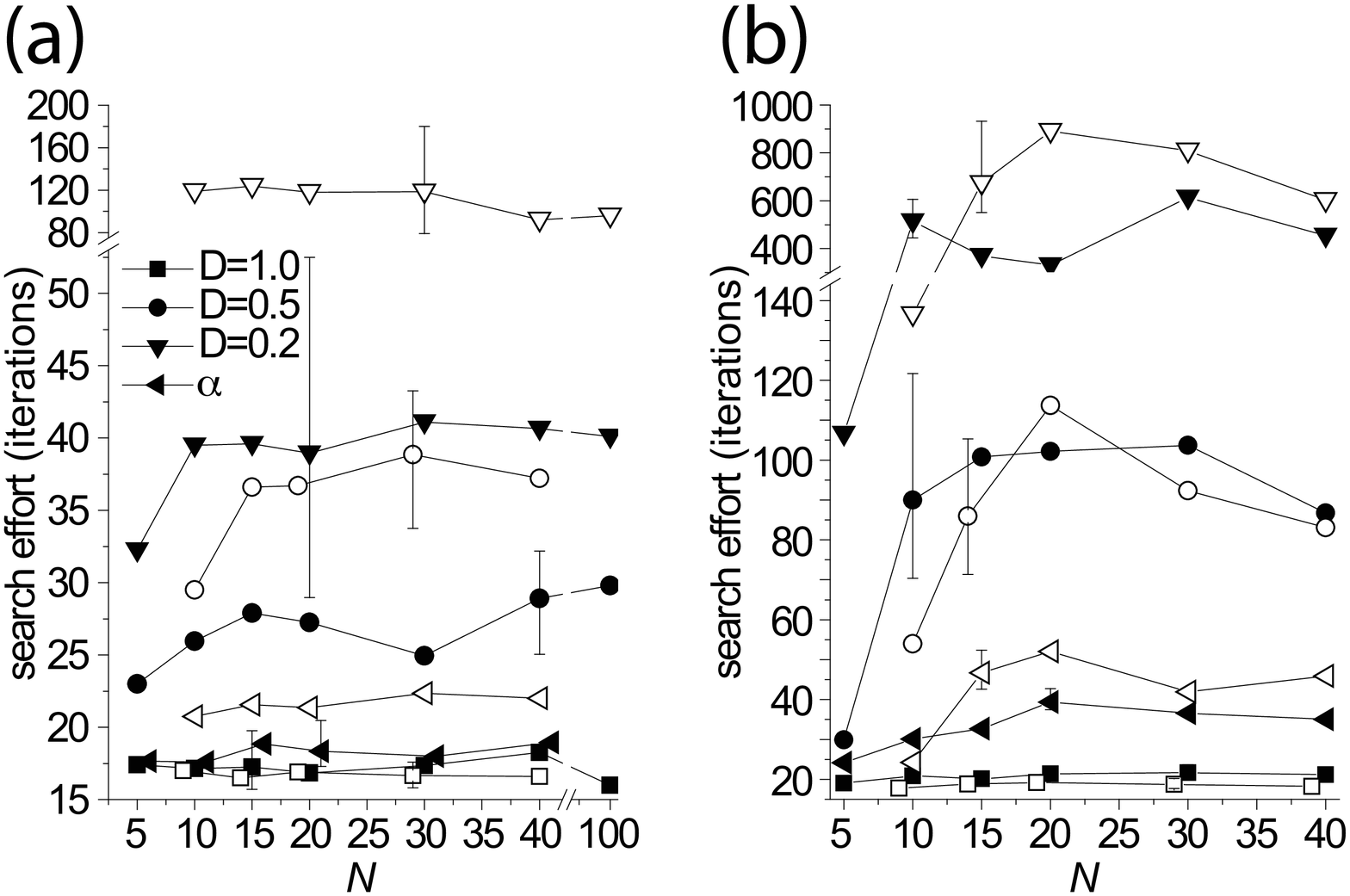}
\caption{ \label{dipscale}}
\end{figure}

Figure \ref{dipscale} shows the mean search effort versus $N$ for rotor $H_0$ (Eq. (\ref{ho}), (a)) and oscillator $H_0$ (Eq. (\ref{osc}), (b)) with $D$=1.0, 0.5, 0.2, and $\alpha$ dipoles and the transitions $|1\rangle\to|5\rangle$ (solid symbols) and $|1\rangle\to|10\rangle$ (open symbols). Representative statistical error bars are presented for one value of $N$ for each choice of $D$ and $P_{i\to f}$. Error bars for other $N$ (with the exception of the smallest $N$ for the oscillator $H_0$ structure) were of similar magnitude. Examination of Figure \ref{dipscale} shows two striking trends. First, the search effort for any choice of dipole structure is invariant to $N$, at least for $N\gtrsim 10$ for the $|1\rangle\to|5\rangle$ transition and $N\gtrsim 15$ for the $|1\rangle\to|10\rangle$ transition. This result agrees with earlier work using kinematic control variables \cite{me}. Second, for the same dipole structure and target transition, the oscillator $H_0$ structure requires a greater effort than for the corresponding conditions with the rotor $H_0$ when the dipole coupling is weak ($D\leq 0.5$). This result shows how the choice of $H_0$ produces landscapes with different local structures, as will be reported in Section \ref{searchstruct}.

\begin{figure}
\includegraphics[width=8.5cm]{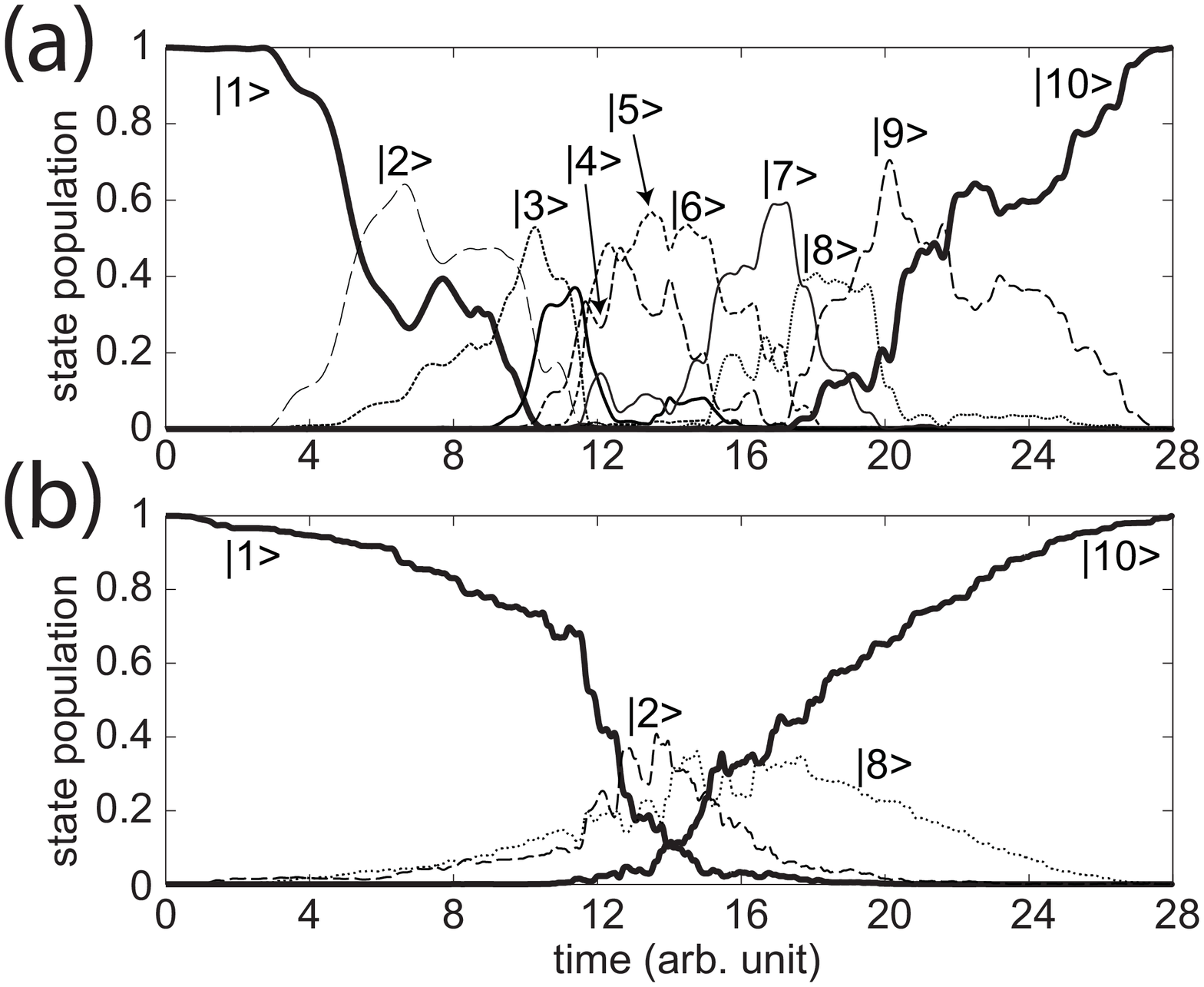}
\caption{ \label{mechanism}}
\end{figure}

A more detailed examination of Figure \ref{dipscale} reveals two further trends. Stronger coupling (i.e., $D$=1 and $\alpha$ dipoles) results in more efficient searches. This intuitive result can be explained in terms of the accessible mechanistic pathways connecting $|i\rangle$ and $|f\rangle$. With strong coupling, both ``ladder climbing'' (i.e., transitions between adjacent states) and quasi-direct transitions are accessible, making it easier to find an optimal field that exploits one of many pathways from $|i\rangle$ to $|f\rangle$. With weak coupling, accessibility of only adjacent transitions limits the number of pathways, thus making it more difficult to find a field that utilizes one of them. This phenomenon is illustrated in Figure \ref{mechanism}, which shows the population of each state $|1\rangle$ through $|10\rangle$ of a 10-level system plotted versus time, with the goal to transfer all population to $|10\rangle$ at $T$=28. In Figure \ref{mechanism}(a) ($D$=0.2), each intermediate state $|2\rangle$ through $|9\rangle$ is accessed sequentially in going from $|1\rangle\to|10\rangle$. All such plots for $D$=0.2 showed involvement of each intermediate state. In Figure \ref{mechanism}(b) ($D$=1.0), only states $|2\rangle$ and $|8\rangle$ are involved; the remaining intermediate states were never populated more than 10$\%$. Other plots for $D$=1.0 showed between one and eight intermediate states involved, indicating more accessible pathways between $|1\rangle$ and $|10\rangle$. Finally, for both $H_0$ structures, the more distant $|1\rangle\to|10\rangle$ transition generally requires more effort than the closer $|1\rangle\to|5\rangle$ transition, except when $D$=1.0, where the effort is similar. This result can be understood in terms of the dipole coupling as well. When $D$=1.0, all transitions are equally allowed, so changing the final target state does not affect the number of accessible mechanistic pathways, resulting in no increase of search effort. Weaker coupling, however, closes off pathways between non-adjacent states, further reducing the number of accessible pathways as the distance between the initial and final states is increased. Plots of state population versus time similar to Figure \ref{mechanism} confirm this behavior (not shown).

\subsection{Varying the Initial Control Field}\label{field}
In order to isolate the effects of varying the initial control field on the required search effort, the rotor $H_0$ (Eq. (\ref{ho})), $\alpha$ dipole structure and the $|1\rangle\to|5\rangle$ transition were fixed. Sets of 20 simulations were performed for initial field strength $F$=0.1, 10, and 100 with a bandwidth bounded by $\Omega$=$\omega_{5}$ in order to determine the effect of initial fluence (i.e., $F^2$) on search effort; the fluence was allowed to vary freely during the landscape ascent. Fields of initial strength $F$=1 with a fixed maximal frequency $\Omega$=$\omega_{20}$ as well as an $N$-dependent maximal frequency $\Omega$=$\omega_{N}/2$ were also chosen in order to determine the effects of providing more bandwidth than necessary.

\begin{figure}
\includegraphics[width=8.5cm]{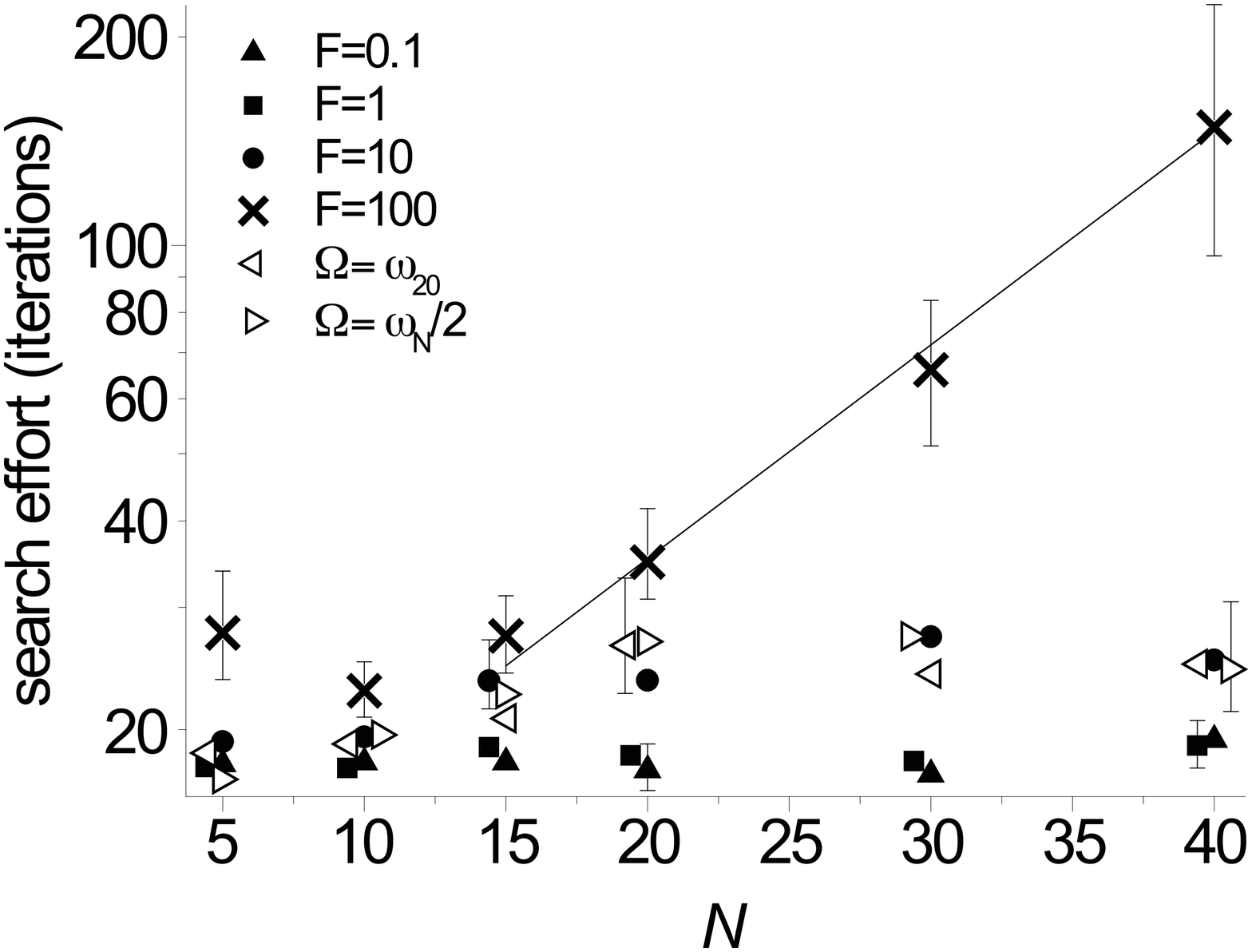}
\caption{\label{fieldscale}}
\end{figure}

\begin{figure}
\includegraphics[width=8.5cm]{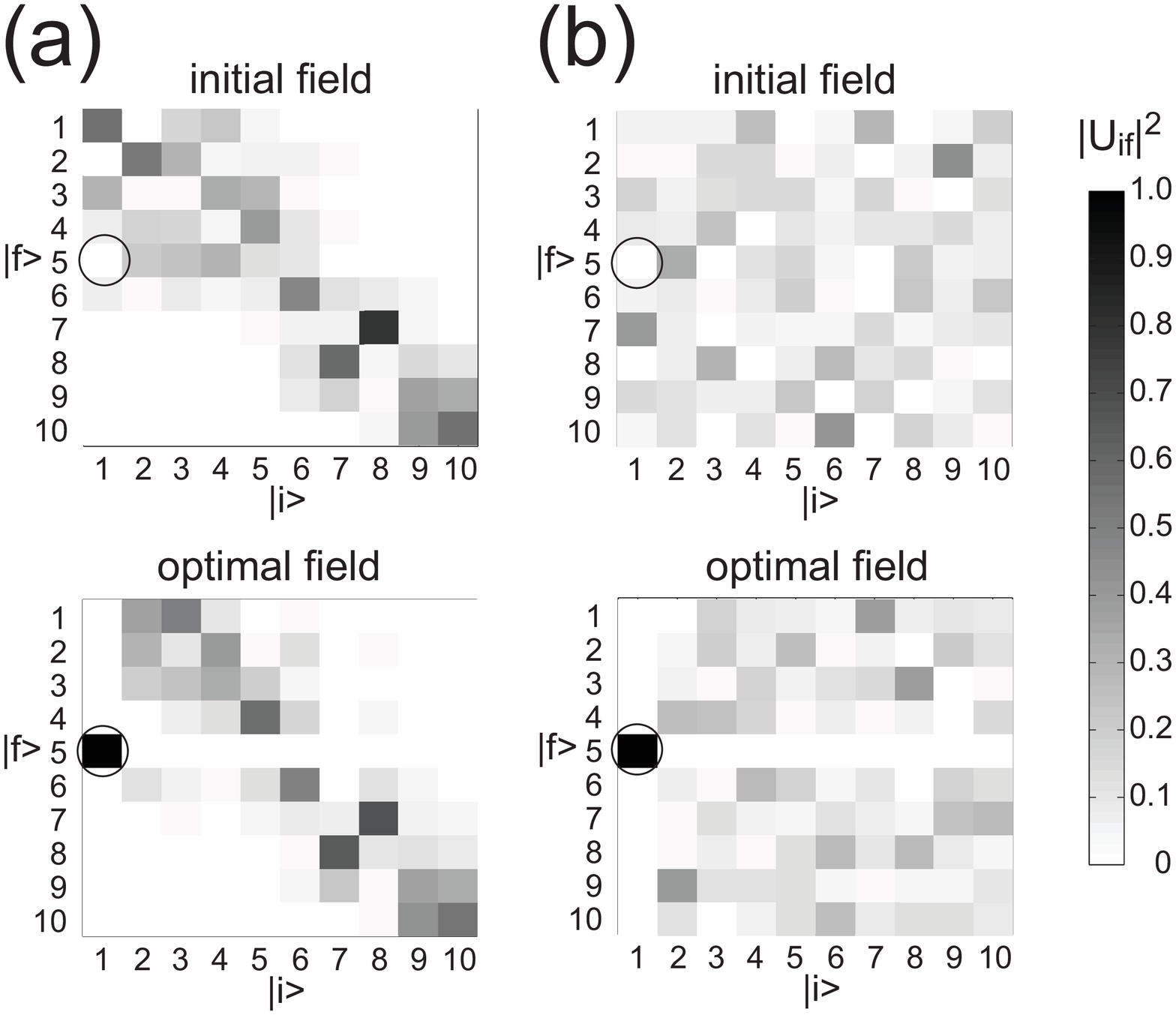}
\caption{\label{us}}
\end{figure}

Figure \ref{fieldscale} presents the mean search effort versus $N$ with representative statistical variation bars. The effort is similar for $F$=0.1 and $F$=1 (included as a reference), and $F$=10; these searches are the most efficient. Further increasing the field strength leads to greater search effort:
at $F$=100, the effort scales exponentially for $N\geq10$ (note the least squares line and the log scale of the ordinate). This result appears to arise because a strong field can easily spread an initial amplitude out among many states, making it difficult to then gather all of the amplitude into the target state $|f\rangle$. This conclusion can be verified by examining the matrix with elements $\{|U_{if}(T,0)|^2\}$ produced by the initial and optimal electric fields. When $F$=1, the initial matrix $\{|U_{if}(T,0)|^2\}$ is nearly diagonal, since the far off-diagonal elements (including the desired (5,1) element) are close to zero, as shown in Figure \ref{us}(a). In contrast, when $F$=100 in Figure \ref{us}(b), the initial matrix $\{|U_{if}(T,0)|^2\}$ contains many significant off-diagonal elements, indicating that the amplitude is spread out through many states. The $\{|U_{if}(T,0)|^2\}$ matrices produced by the optimal fields retain the predominantly diagonal structure for $F$=1 and the significant off-diagonal elements for $F$=100 (Figure \ref{us}, bottom row). When the bandwidth provided is more than necessary to make the $|1\rangle\to|5\rangle$ transition, the effort grows very slowly with $N$. The slight increase in effort compared to using the maximal frequency $\Omega$=$\omega_{5}$ suggests that additional access to unneeded ancillary states makes it more difficult to gather all of the amplitude in the target final state; examination of the $\{|U_{if}(T,0)|^2\}$ matrices for these cases verified this behavior (not shown).

\subsection{The $|1\rangle\to|N\rangle$ Transition}\label{1n}
The simulations above employed a fixed choice of $|i\rangle$ and $|f\rangle$ as $N$ was increased. Specifying $|1\rangle\to|N\rangle$ as the target transition causes the final state to recede from the initial state as $N$ is increased. To accommodate the increasing demands of transferring amplitude between successively more distant states, the strength of the initial fields was chosen as $F$=10 and the frequencies were chosen on an $N$ dependent bandwidth with maximal frequency $\Omega$=$\omega_{N}/2$. Because the initial population in $|N\rangle$ drastically decreases with rising $N$ (c.f., Figure \ref{stats}), iterations were counted starting when the yield reached $P_{i\to f}$=0.001 to normalize the effort against this discrepancy. 

\begin{figure}
\includegraphics[width=8.5cm]{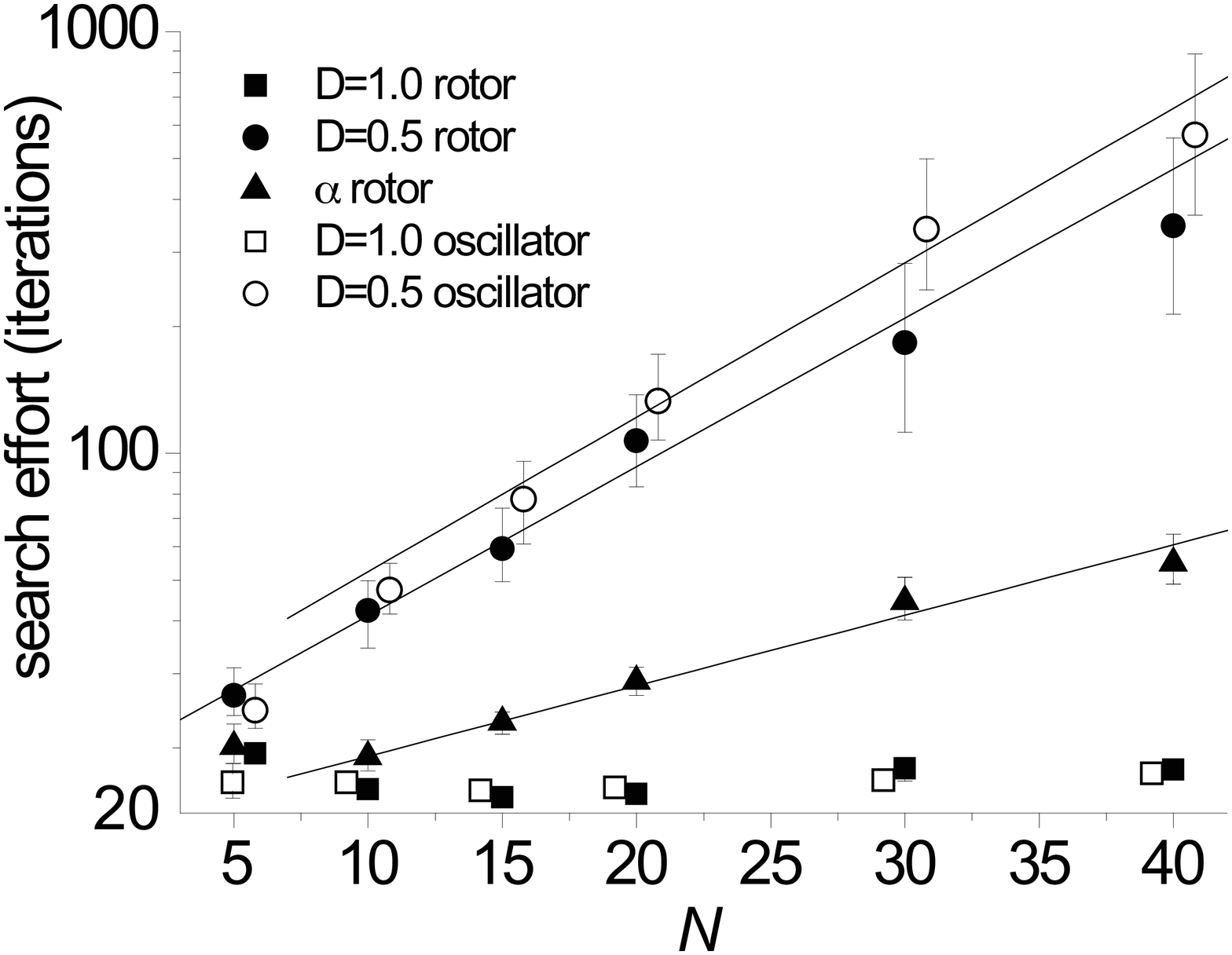}
\caption{\label{1nscale}}
\end{figure}

The results of simulations using $D$=1.0, 0.5 and $\alpha$ dipole structures with both rotor and oscillator $H_0$ structures are shown in Figure \ref{1nscale}. The scaling behavior with $N$ changes significantly depending on the dipole structure. When $D$=1.0, the effort is invariant to $N$. Although the distance between the initial and final states is rising with $N$, when all transitions are equally allowed, the number of possible pathways between $|1\rangle$ and $|N\rangle$ is large enough to permit efficient optimization even at large $N$. In contrast, for the $\alpha$ and $D$=0.5 dipoles, the effort scales exponentially with $N$, as shown with the least squares fit lines on the semi-log plot in Figure \ref{1nscale}. The 12 falsely trapped cases mentioned in Section \ref{notraps} were for these simulations employing 2048 time-points with $D$=0.5 and $N$=30 and 40. The additional resolution gained upon interpolation of the control field on 4096 time-points eliminated these false traps with further climbing iterations. These iterations were added for computation of the mean search effort in Figure \ref{1nscale}. Receding target objectives with increasing system complexity (i.e., illustrated here with $|1\rangle\to|N\rangle$) are generally not the case for laboratory OCE, thereby evidently avoiding the exponential scaling of effort.

The observed systematic invariance of search effort with respect to $N$ over a wide range of Hamiltonian and initial control field structures verifies that the search effort for population transfer does not depend on the system complexity, as was the case for kinematic controls \cite{me}. This result is valid upon making a rational choice of the control objective and initial field (i.e., for fixed target transition and reasonable initial field strength). The results suggest that under such circumstances, controlling complex quantum systems with many degrees of freedom should be no more difficult than controlling simple systems. Evidently the same conclusion applies to performing OCE for various objectives, where the search effort appears to be essentially the same regardless of the system complexity when operating with physically appropriate controls \cite{wen,Bartels2000,herek,Miller06}. The next section will address the relationship between the observed trends in search effort and the underlying control landscape structure. 

\section{Search Effort and Landscape Structure}\label{searchstruct}

Examination of the relationship between the structure of the control landscape and the required search effort makes it possible to obtain further insight into the scaling results obtained in Section \ref{Scale}. In this section, we determine the local landscape structure in terms of the metrics defined in Section \ref{struct}. Here, the notion of structure refers to landscape features other than topological critical points; the landscape theory predicts critical points only at $P_{i\to f}$=0 and 1, which was verified by the observed lack of traps in Section \ref{notraps}.

\subsection{Search Trajectories on the Control Landscape}\label{dist}
We first consider the relationship between the search effort and the complexity of the trajectories over the landscape taken during the optimal searches using the ratio metric $R_{\varepsilon}$ defined in Eq. (\ref{ratio}). The mean values of $R_{\varepsilon}$ were calculated for all the searches performed in Section \ref{Scale}. Select examples with the rotor $H_0$ structure (Eq. (\ref{ho})) are plotted in Figure \ref{path}. When the search effort is invariant to $N$ (i.e., the $|1\rangle\to|5\rangle$ transition with $F$=1 for $\alpha$ and $D$=0.2 dipoles in Figure \ref{path}), the ratio $R_{\varepsilon}$ is also invariant to $N$, in agreement with kinematic results \cite{me}. In contrast, when effort increases with $N$ (e.g., the $|1\rangle\to|N\rangle$ transition or large strength $F$), the path length correspondingly rises with $N$. For all conditions where search effort is invariant to $N$, the ratio $R_{\varepsilon}$ is correlated to the search effort, as shown in Table 1 for simulations using the rotor $H_0$ (left of double line) and oscillator $H_0$ (right of double line); ratios are significantly higher for the oscillator $H_0$, although these do not scale with $N$. The values of the distances $||\Delta_E \varepsilon||$ and $||\Delta_P \varepsilon||$ used to define $R_{\varepsilon}$ follow the same correlations with effort. The differences in values of $R_{\varepsilon}$ between optimizations using the rotor and oscillator $H_0$ structures for weakly coupled dipoles can be explained by examination of the landscape slope, as discussed below.

\subsection{Landscape Slopes and Search Effort}\label{slopes}
The magnitude of the gradient $\mathcal{S}_m$ provides valuable information about how fast a search algorithm may improve the yield. Intuitively, a steep slope would be conducive to efficient optimization because the yield may improve rapidly upon taking an algorithmic step, while a very shallow slope should slow the optimization.

For the optimizations in Section \ref{Scale}, the slope metric $\mathcal{S}_0$ at the initial random control field (or at the first iteration where $P_{i\to f}\geq$0.001) and the point of maximal slope $\mathcal{S}_{max}$ were recorded; at $P_{i\to f}\sim$0.001, the slope metric $\mathcal{S}_0$ is typically small. Both the initial $\mathcal{S}_0$ and maximal $\mathcal{S}_{max}$ slope metrics along an optimization may be expected to correlate with the required search effort. Figure \ref{slope} shows the mean value of the initial slope metric $\mathcal{S}_0$ (filled symbols) and maximal slope metric $\mathcal{S}_{max}$ (open symbols) for selected optimizations from Section \ref{Scale}. The initial and final slope metrics are independent of $N$ under conditions where the search effort is also invariant, while both metrics for the $|1\rangle\to|N\rangle$ transition decrease as $N$ rises, in accordance with the increase in search effort. All conditions where the effort was dependent on $N$ exhibited the behavior of decreasing slope metrics as $N$ rises. For the cases invariant to $N$, more difficult optimizations (e.g., optimization with a weak dipole) have smaller initial and maximal slope metrics than easier optimizations, as shown in Table 1. Thus, the search effort follows the intuitive conjecture that a steeper slope results in more efficient optimization, as was found using kinematic control variables \cite{me}. In general, the linkage of search effort to the gradient depends on the choice of search algorithm. Most OCT studies use gradient algorithms, so in such cases the search effort may be expected to depend on the initial and/or maximal slope metric. However, other ``smart'' algorithms (e.g., with stochastic logic) can also exploit the favorable slopes and direct pathways to the optimum with $R_{\varepsilon}$ being small.

An exception to the simple search effort correlation with the initial and maximal slope metrics arises for searches using weakly coupled dipoles when comparing the two $H_0$ structures with otherwise identical search conditions. The effort for the oscillator $H_0$ is drastically higher than for the rotor $H_0$, but the initial and maximal slope metrics are of similar magnitude, as shown in Figure \ref{slope} and Table 1. This discrepancy can be explained by examining the trajectory of the slope metric and the ratio $R_\varepsilon$ over the course of an optimization. As an example, these trajectories for searches with $N$=20, $D$=0.2 and $|1\rangle\to|5\rangle$ transition are compared for the two different $H_0$ structures. Figure \ref{traject} shows the trajectory of the slope metric (a) and the trajectory of $R_\varepsilon$ (b) for two searches with each $H_0$ structure. The trajectories of the slope metric $\mathcal{S}$ for the rotor $H_0$ share the simple structure of starting near zero at the initial field with $P_{i\to f}\sim0.001$, rising to a maximum around $P_{i\to f}\sim0.5$, and decreasing towards the optimum. Similarly, the trajectories of $R_\varepsilon$ for these searches show a simple monotonic rise with $P_{i\to f}$. In contrast, the trajectories of searches using the oscillator $H_0$ structure show a more complex behavior over the landscape. Instead of reaching a high at $P_{i\to f}\sim0.5$, the maximal slope metric for the oscillator searches occurs below $P_{i\to f}\sim0.3$, and the slope decreases rapidly thereafter. Examination of $R_\varepsilon$ at $P_{i\to f}$ values (b) corresponding to the rapidly decreasing slope metric in (a) shows a fast jump in $R_\varepsilon$ with $P_{i\to f}$, indicating a relatively ``gnarled'' landscape region. Finally, the slope metric for the oscillator searches drops quickly for $P_{i\to f}\geq 0.8$, and the ratio $R_\varepsilon$ rises accordingly. Other trajectories for searches using the oscillator $H_0$ with a weakly-coupled dipole show similar features, suggesting that an oscillator $H_0$ structure with a weakly coupled dipole inherently creates a more gnarled landscape than a rotor $H_0$ with the same dipole.

\subsection{Second Order Landscape Structure}\label{o2}
Examination of the second-order landscape structure metrics can provide further insight into contributions to the relative search effort required under different optimization conditions. Calculations of the Hessian matrix and associated structure metrics at the bottom and top of the landscape were performed on the rotor $H_0$ structure for (i) the $|1\rangle\to|5\rangle$ transition with the $\alpha$ and $D$=0.5 dipole structures for $F$=1, (ii) the $\alpha$ dipole structure for $F$=100, and (iii) $D$=0.5 for the $|1\rangle\to|N\rangle$ transition with $F$=10. With the oscillator $H_0$ structure, the calculations were performed for the $D$=0.5 dipole and $F$=1 with $|1\rangle\to|5\rangle$ transition. In order to obtain Hessian matrices reliably representing the bottom and top of the landscape, all optimizations began at $P_{i\to f}\leq1\times10^{-5}$ and the convergence criterion was $P_{i\to f}\geq0.99999$.

It has been shown theoretically that the Hessian spectrum at the bottom of the landscape has at most two nonzero positive eigenvalues and the spectrum at the top contains at most 2$N$-2 nonzero negative eigenvalues \cite{demiralp}. This analysis is verified by our numerical results. Figure \ref{eigs} shows the Hessian spectra at the top of the landscape for individual optimizations of $|1\rangle\to|5\rangle$ transition with rotor $H_0$ structure, $F$=1, and $D$=1.0 (Figure \ref{eigs}(a), for $N$ ranging from 5 to 30) and $D$=0.5 (Figure \ref{eigs}(b), for $N$ ranging from 5 to 15). The vertical dotted lines denote the eigenvalue index of 2$N$-2 for each $N$ reported.  In the case of $D$=1.0, there is always a clear distinction between the (2$N$-2)th eigenvalue ($\sim-10$) and the (2$N$-3)th eigenvalue ($\gtrsim -0.01$). The magnitude of the largest and smallest nonzero eigenvalues does not change with $N$. For $D$=0.5, the drop in eigenvalue magnitude at the index 2$N$-2 is apparent at $N$=5 and 10 (note log scale on the ordinate in Figure \ref{eigs}(b)). By $N$=15, the distinction between the final nonzero and first zero eigenvalue is expected to occur between the 28th and 29th eigenvalues, however the eigenvalues are already of very small magnitude by the 23rd eigenvalue. Recording the eigenvalues for larger values of $N$ with $D$=0.5 revealed similar patterns of eigenvalue behavior. This result shows that for large $N$ with weak dipole couplings, fewer than 2$N$-2 negative eigenvalues can be expected at the top of the landscape, and there is no clear boundary between the zero and nonzero eigenvalues. Fewer than 2$N$-2 nonzero eigenvalues were also observed for $N\geq 15$ using the oscillator $H_0$ structure with $D$=0.5 (not shown). With strong dipolar couplings (i.e., $D$=1.0), there are always exactly 2$N$-2 nonzero eigenvalues; for the $\alpha$ dipole structure, exactly 2$N$-2 eigenvalues persist through $N$=30, and by $N$=40 there are fewer than 2$N$-2 eigenvalues (not shown). At the bottom of the landscape, there is a clear distinction between the two positive eigenvalues and the remaining zero eigenvalues, which occurred under all search conditions (not shown). These observations about the Hessian eigenvalues at the bottom and top of the landscape validate the theoretically predicted spectra \cite{demiralp}. Additionally, the number of non-zero Hessian eigenvalues at the top of the landscape influences the robustness of the control outcome to field noise; the presence of fewer such eigenvalues enhances the robustness \cite{vinny2}.

Examination of the Hessian trace and curvature metrics (c.f. Eqs. (\ref{trh}) and (\ref{curve})) at the bottom and top of the landscape yielded intuitive correlations between these metrics and the required search effort, as was the case with the slope metric. As graphs of these metrics versus $N$ are similar to Figure \ref{slope}, the data are not plotted again. Near the bottom of the landscape, both the Hessian trace and curvature metrics are invariant to $N$ when the search effort is also invariant, and smaller values of these metrics are recorded for more difficult search conditions (e.g., oscillator $H_0$, small dipole coupling). Where exponential scaling of search effort with $N$ was found, both metrics decrease exponentially with $N$ near the bottom of the landscape. At the top of the landscape ($P_{i\to f}\geq0.99999$), the Hessian trace is proportional to $N$, regardless of search parameters, due to its dependence on the dipole norm $||\mu||^2$ \cite{demiralp}. The curvature exhibits intuitive correlation with the search effort, remaining constant with $N$ for cases that lack search effort scaling, and decreasing in magnitude with $N$ where scaling is observed. Thus, all of the landscape structure metrics examined in this section correlate in an intuitive way with the required optimization search effort. These results show that the landscape structure metrics provide a good method to predict the relative required search effort under a variety of conditions.

\section{Conclusion}\label{conclusion}

This work addressed two major issues surrounding optimal control of population transfer in quantum systems. The first objective explored the fundamental topic of whether suboptimal trapping extrema are encountered while searching for an optimal control field. The second objective examined how the required effort to find an optimal control field scales with the complexity of the quantum system as measured by its size $N$. 

The possible existence of traps on the control landscape is of both basic and practical importance. Quantum control landscapes can rigorously be shown to contain no traps under simple physical assumptions \cite{mike1, demiralp, shen, taksan}. The vast OCT literature supports the ability to reach excellent yields \cite{shi,zhu,judson,kral,sola,grossmann,RabO3,abe,artamonov,artamonov2,kaluza,gross,nakagami,abe1,lapert,salomon,zou,ambrosek,amstrup,balintkurti,botina,jakubetz,krause,turinici,ohtsuki,ohtsuki2,ohtsuki3,ohtsuki4,phan,ren,schirmer,shah,somloi,wang,wang2,zhu2,zhu3,zhu4}, although these works are not definitive with regard to the landscape due to control field constraints typically being present. The recent identification of trapping conditions \cite{schirmer2,tannor2,rebing} under unusual circumstances necessitates a more explicit investigation of whether traps can be expected when performing normal optimizations.

{\it The simulations in this work found no evidence of trapping behavior on the control landscape for $P_{i\to f}$.} Of the $\sim$5000 searches performed, a total of 12 were initially found to be putative traps warranting further investigation. Enhancing the time resolution established that the latter traps were in fact false, with all optimization searches then reaching $P_{i\to f}>0.999$. The identification of false trapping behavior due to numerical constraints illustrates the need for special care in performing simulations and the general need for due attention to all physical constraints on the field dynamics when a high yield is desired. The lack of observed traps on the $P_{i\to f}$ landscape is consistent with results reported for the landscape corresponding to the generation of arbitrary unitary transformations $U(T,0)$, where $\sim$20,000 optimizations were performed, all of which reached an optimal fidelity value \cite{mew}.

The second issue studied here of search effort scaling with $N$ is primarily of practical importance, indicating whether the control of large, complex quantum systems in the laboratory is feasible. The OCT literature collectively suggests that the required search effort to find an optimal control may be independent of the complexity (i.e., here captured by $N$) of the target quantum system \cite{zhu,judson,RabO3,kaluza,nakagami,lapert,ambrosek,amstrup,balintkurti,botina,jakubetz,krause,turinici,ohtsuki,ohtsuki2,ohtsuki3,ohtsuki4,phan,ren,schirmer,shah,somloi,wang,wang2,zhu2,zhu3,zhu4}. The results from this work systematically verify this behavior and identify the control conditions sufficient for the search effort scaling to be independent of $N$. Specifically, choosing a fixed target transition $|i\rangle\to|f\rangle$ results in the scaling of effort being invariant to $N$ across a wide range of dipole matrix structures and reasonable initial control field parameters, although the absolute search effort can vary widely. This attractive behavior breaks down, however, upon choosing targets that themselves increase in complexity with the system (e.g., $|1\rangle\to|N\rangle$) or starting with a large initial control field strength for a fixed target transition, where the wavefunction amplitude spreads widely before finally being drawn into the target state. 

The observed search effort was found to correlate with the landscape features, as measured by the distance and structure metrics. The scaling of the ratio of path length to Euclidian distance $R_{\varepsilon}$ with $N$ follows that of the search effort; $R_{\varepsilon}$ only increases with $N$ for the difficult cases such as the $|1\rangle\to|N\rangle$ target or with a large initial field fluence. For cases with scaling invariant to $N$, the relative search effort can be predicted by the value of $R_{\varepsilon}$, with greater values of this metric correlating with a greater search effort. Analysis of the local structure of the landscape shows that the search effort correlates with the slope metric (gradient norm) in an intuitive manner. A steeper landscape slope both at the initial control field and at the point of maximal slope results in a lower search effort than a shallow slope. The landscape slopes at these points are invariant to $N$, except for the cases where the search effort scales with $N$, for which both initial and maximal slopes decrease as $N$ rises. A similar correlation of search effort with the curvature metric near the bottom and top of the landscape with $N$ was observed. Finally, the collective $dynamic$ findings on search effort show a strong relation to analogous behavior found using $kinematic$ variables \cite{me}. Although clearly additional dynamical features occur (e.g., through the amplitude and structure of the dipole couplings), much of the basic invariant scaling findings with $N$ appear to have their origins in the underlying simpler kinematic control formulation. 

This work addressed many classes of control Hamiltonians in order to demonstrate the broad applicability of the two main results in this work. However, some classes of quantum systems, such as those containing degenerate energy levels or additional symmetry in the dipole matrix, were not addressed here. Provided that such systems are controllable \cite{ramakrishna} (e.g., where dipole couplings break the symmetry produced by degenerate states), the favorable topological and scaling results are expected to hold. For other special classes of systems that are uncontrollable or nearly so (e.g., a harmonic oscillator), special care in the choice of controls may be needed to avoid traps on the landscape arising from the lack of system controllability. Most classes of quantum systems, however, are expected to satisfy the controllability requirement and thus exhibit qualitatively similar behavior in terms of landscape topology and search effort seen here.

The favorable scaling of $P_{i\to f}$ with $N$ suggests that optimization of state preparation with a suitable set of controls should be relatively easy to attain using OCE, even with complex systems. Although the quantum systems employed here do not model any particular real system, the results using the rigid rotor and anharmonic oscillator $H_0$ structures indicate that some quantum systems may generate a landscape with a more gnarled local structure than others, leading to wide variations in the absolute search effort required to find an optimal control. Nevertheless, a family of quantum systems that are difficult to optimize may still show invariant scaling with $N$. These results are consistent with successful OCE studies on complex molecules such as proteins \cite{herek,Miller06}, even though the laboratory conditions are more involved than the ideal circumstances presented here.

Overall, this work demonstrated that both the topology and the local structure of the control landscape for population transfer are conducive to efficient optimal control. Extensive simulations did not encounter traps on the landscape upon reasonable choices of Hamiltonians, initial control fields, and careful numerical optimization. The invariance of scaling of the search effort with system complexity was shown to be due to favorable local landscape structure that does not grow more complex with system size $N$. Besides state preparation, recent studies generalize these landscape topology, features, and optimization scaling results to the preparation of unitary transformations \cite{mew} and broader classes of observables \cite{greg}.

{\bf Acknowledgment.} The authors acknowledge support from Department of Energy grant DE-FG02-02ER15344. K.W.M. acknowledges the support of a NSF graduate student fellowship.

\bibliography{theorypapers}

\begin{thebibliography}{10}

\bibitem{shi}
Shi, S. and Rabitz, H.
\newblock {\em J. Chem. Phys.}{ \bf 92}, 364 (1990).

\bibitem{zhu}
Zhu, W., Botina, J., and Rabitz, H.
\newblock {\em J. Chem. Phys.}{ \bf 108}, 1953 (1998).

\bibitem{judson}
Judson, R. and Rabitz, H.
\newblock {\em Phys. Rev. Lett.}{ \bf 68}, 1500 (1992).

\bibitem{constantin}
Brif, C., Chakrabarti, R., and Rabitz, H.
\newblock {\em {New J. Phys.}}{ \bf {12}}, {075008} ({2010}).

\bibitem{raj}
Chakrabarti, R. and Rabitz, H.
\newblock {\em {Int. Rev. Phys. Chem.}}{ \bf {26}}({4}), {671--735} ({2007}).

\bibitem{kral}
Kral, P., Thanopulos, I., and Shapiro, M.
\newblock {\em Phys. Rev. A}{ \bf 72}, 020303 (2005).

\bibitem{sola}
Sola, I., Malinovsky, V.~S., and Tannor, D.~J.
\newblock {\em Phys. Rev. A}{ \bf 60}, 3081 (1999).

\bibitem{grossmann}
Grossman, F., Feng, L., Schmidt, G., Kunert, T., and Schmidt, R.
\newblock {\em Europhysics Letters}{ \bf 60}, 201 (2002).

\bibitem{RabO3}
Artamonov, M. and Rabitz, H.
\newblock {\em Chem. Phys.}{ \bf 305}, 213 (2004).

\bibitem{abe}
Abe, M., Ohtsuki, Y., Fujimura, Y., Lan, Z., and Domcke, W.
\newblock {\em J. Chem. Phys.}{ \bf 124}, 224316 (2006).

\bibitem{artamonov}
Artamonov, M., Ho, T., and Rabitz, H.
\newblock {\em J. Chem. Phys.}{ \bf 124}, 064306 (2006).

\bibitem{artamonov2}
Artamonov, M., Ho, T., and Rabitz, H.
\newblock {\em Chem. Phys.}{ \bf 328}, 147 (2006).

\bibitem{kaluza}
Kaluza, M., Muckerman, J., Gross, P., and Rabitz, H.
\newblock {\em J. Chem. Phys.}{ \bf 100}, 4211 (1994).

\bibitem{gross}
Gross, P., Bairagi, D., Mishra, M., and Rabitz, H.
\newblock {\em Chem. Phys.}{ \bf 223}, 263 (1994).

\bibitem{nakagami}
Nakagami, K., Ohtsuki, Y., and Fujimura, Y.
\newblock {\em J. Chem. Phys.}{ \bf 117}, 6429 (2002).

\bibitem{abe1}
Abe, M., Ohtsuki, Y., Fujimura, Y., and Domcke, W.
\newblock {\em J. Chem. Phys.}{ \bf 123}, 144508 (2005).

\bibitem{lapert}
Lapert, M., Tehini, R., Turinici, G., and Sugny, D.
\newblock {\em Phys. Rev. A}{ \bf 78}, 023408 (2008).

\bibitem{salomon}
Salomon, J., Dion, C.~M., and Turinici, G.
\newblock {\em J. Chem. Phys.}{ \bf 123}, 144310 (2005).

\bibitem{zou}
Zou, S., Balint-Kurti, G.~G., and Manby, F.~R.
\newblock {\em J. Chem. Phys.}{ \bf 127}, 044107 (2008).

\bibitem{wen}
Wen, H., Rangan, C., and Bucksbaum, P.
\newblock {\em Phys. Rev. A}{ \bf 68}, 053405 (2003).

\bibitem{Brixner}
Brixner, T., Damrauer, N.~H., Kiefer, B., and Gerber, G.
\newblock {\em J. Chem. Phys.}{ \bf 118}, 3692 (2003).

\bibitem{Baumert}
Assion, A., Baumert, T., Bergt, M., Brixner, T., Kiefer, B., Seyfried, V.,
  Strehle, M., and Gerber, G.
\newblock {\em Science}{ \bf 282}, 919 (1998).

\bibitem{Levis2001}
Levis, R., Menkir, G., and Rabitz, H.
\newblock {\em Science}{ \bf 292}, 709 (2001).

\bibitem{Gerber02}
Bergt, M., Brixner, T., Dietl, C., Kiefer, B., and Gerber, G.
\newblock {\em J. Organomet. Chem.}{ \bf 661}, 199 (2002).

\bibitem{Bartels2000}
Bartels, R., Backus, S., Zeek, E., Misoguti, L., Vdovin, G., Christov, I.~P.,
  Murnane, M.~M., and Kapteyn, H.~C.
\newblock {\em Nature}{ \bf 406}, 164 (2000).

\bibitem{Bartels2004}
Bartels, R., Murnane, M., Kapteyn, H., Christov, L., and Rabitz, H.
\newblock {\em Phys. Rev. A}{ \bf 70}, 1 (2004).

\bibitem{Pfeifer2005}
Pfeifer, T., Kemmer, R., Spitzenfeil, R., Walter, D., Winterfeldt, C., Gerber,
  G., and Spielmann, C.
\newblock {\em Opt. Lett.}{ \bf 30}, 1497 (2005).

\bibitem{herek}
Herek, J.~L., Wohlleben, W., Cogdell, R.~J., Zeidler, D., and Motzkus, M.
\newblock {\em Nature}{ \bf 417}, 533 (2002).

\bibitem{Gerber05}
Vogt, G., Krampert, G., Niklaus, P., Neurnberger, P., and G., G.
\newblock {\em Phys. Rev. Lett.}{ \bf 94}, 068305 (2005).

\bibitem{Miller06}
Prokhorenko, V.~I., Nagy, A.~M., Waschuk, S.~A., Brown, L.~S., Birge, R.~R.,
  and Miller, R. J.~D.
\newblock {\em Science}{ \bf 313}, 1257--1261 (2006).

\bibitem{mike1}
Rabitz, H., Hsieh, M., and Rosenthal, C.
\newblock {\em Science}{ \bf 303}, 1998--2001 (2004).

\bibitem{ramakrishna}
Ramakrishna, V., Salapaka, M.~V., Dahleh, M., Rabitz, H., and Pierce, A.
\newblock {\em Phys. Rev. A}{ \bf 51}, 960 (1995).

\bibitem{demiralp}
Rabitz, H., Ho, T.-S., Hsieh, M., Kosut, R., and Demiralp, M.
\newblock {\em Phys. Rev. A}{ \bf 74}, 012721 (2006).

\bibitem{shen}
Shen, Z., Hsieh, M., and Rabitz, H.
\newblock {\em J. Chem. Phys.}{ \bf 124}, 204106 (2006).

\bibitem{taksan}
Ho, T. and Rabitz, H.
\newblock {\em J. Photo. Chem. A}{ \bf 180}, 226 (2006).

\bibitem{schirmer2}
de~Fouquieres, P. and Schirmer, S.~G.
\newblock { \bf } (2011).
\newblock Preprint: arXiv:1004.3492.

\bibitem{tannor2}
Pechen, A.~N. and Tannor, D.~J.
\newblock {\em Phys. Rev. Lett.}{ \bf 106}(12), 120402 (2011).

\bibitem{rebing}
Wu, R., Dominy, J., Ho, T.-S., and Rabitz, H.
\newblock { \bf } (2011).
\newblock Preprint arXiv:0907.2354.

\bibitem{ambrosek}
Ambrosek, D., Oppel, M., Gonzalez, L., and May, V.
\newblock {\em Opt. Commun.}{ \bf 264}, 502 (2006).

\bibitem{amstrup}
Amstrup, B., T\`{o}th, G.~J., Szabo, G., Rabitz, H., and Lorincz, A.
\newblock {\em J. Phys. Chem.}{ \bf 99}, 5206 (1995).

\bibitem{balintkurti}
Balint-Kurti, G., Manby, F.~R., Ren, H., Artamonov, M., Ho, T., and Rabitz, H.
\newblock {\em J. Chem. Phys.}{ \bf 122}, 084110 (2005).

\bibitem{botina}
Botina, J. and Rabitz, H.
\newblock {\em J. Chem. Phys.}{ \bf 104}, 4031 (1996).

\bibitem{jakubetz}
Jakubetz, W., Kades, E., and Manz, J.
\newblock {\em J. Phys. Chem.}{ \bf 97}, 12609 (1993).

\bibitem{krause}
Krause, J., Messina, M., Wilson, K.~R., and Yan, Y.
\newblock {\em J. Phys. Chem.}{ \bf 99}, 13736 (1995).

\bibitem{turinici}
Maday, Y. and Turinici, G.
\newblock {\em J. Chem. Phys.}{ \bf 118}, 8191 (2003).

\bibitem{ohtsuki}
Ohtsuki, Y., Zhu, W.~S., and Rabitz, H.
\newblock {\em J. Chem. Phys.}{ \bf 110}, 9825 (1999).

\bibitem{ohtsuki2}
Ohtsuki, Y., Ohara, N., and Fujimura, Y.
\newblock { \bf 369}, 525 (2003).

\bibitem{ohtsuki3}
Ohtsuki, Y., Turinici, G., and Rabitz, H.
\newblock {\em J. Chem. Phys.}{ \bf 120}, 5509 (2004).

\bibitem{ohtsuki4}
Ohtsuki, Y., Nakagami, K., Fujimura, Y., Zhu, W., and Rabitz, H.
\newblock {\em J. Chem. Phys.}{ \bf 114}, 8867 (2001).

\bibitem{phan}
Phan, M.~Q. and Rabitz, H.
\newblock {\em J. Chem. Phys.}{ \bf 110}, 34 (1999).

\bibitem{ren}
Ren, Q., Balint-Kurti, G.~G., Manby, F.~R., Artamonov, M., Ho, T.-S., and
  Rabitz, H.
\newblock {\em J. Chem. Phys.}{ \bf 124}, 014111 (2006).

\bibitem{schirmer}
Schirmer, S.~G., Girardeau, M.~D., and Leahy, J.~V.
\newblock {\em Phys. Rev. A}{ \bf 61}, 012010 (2000).

\bibitem{shah}
Shah, S.~P. and Rice, S.~A.
\newblock {\em Faraday Discuss.}{ \bf 113}, 319 (1999).

\bibitem{somloi}
Somloi, J., Kazakov, V.~A., and Tannor, D.~J.
\newblock {\em Chem. Phys.}{ \bf 172}, 85 (1993).

\bibitem{wang}
Wang, L., Meyer, H.-D., and May, V.
\newblock {\em J. Chem. Phys.}{ \bf 125}, 014102 (2006).

\bibitem{wang2}
Wang, L. and May, V.
\newblock {\em J. Chem. Phys.}{ \bf 121}, 8039 (2004).

\bibitem{zhu2}
Zhu, W. and Rabitz, H.
\newblock {\em J. Chem. Phys.}{ \bf 109}, 385 (1998).

\bibitem{zhu3}
Zhu, W. and Rabitz, H.
\newblock {\em Phys. Rev. A}{ \bf 58}, 4741 (1998).

\bibitem{zhu4}
Zhu, W. and Rabitz, H.
\newblock {\em Intl. J. Quantum Chem.}{ \bf 93}, 50 (2003).

\bibitem{tannor}
Tannor, D.~J., Kazakov, V., and Orlov, V.
\newblock {\em Time Dependent Quantum Molecular Dynamics}, volume 299 of {\em
  NATO Advanced Study Institute, Series B: Physics}.
\newblock Plenum, New York (1992).

\bibitem{me}
Moore, K., Hsieh, M., and Rabitz, H.
\newblock {\em J. Chem. Phys.}{ \bf 128}, 154117 (2008).

\bibitem{altafini}
Altafini, C.
\newblock {\em J. Math. Phys.}{ \bf 43}, 2051 (2002).

\bibitem{vinny2}
Beltrani, V., Dominy, J., Ho, T.-S., and Rabitz, H.
\newblock {\em J. Chem. Phys.}{ \bf 134}, 194106 (2011).

\bibitem{matlab}
MathWorks, MATLAB, The MathWorks, Natick, MA, 1994.

\bibitem{mike3}
Hsieh, M., Wu, R., Rosenthal, C., and Rabitz, H.
\newblock {\em J. Phys. B: At. Mol. Opt. Phys.}{ \bf 41}, 074020 (2008).

\bibitem{MikeRebing}
Wu, R., Hsieh, M., Rosenthal, C., and Rabitz, H.
\newblock {\em J. Phys. A: Math. Theor.}{ \bf 41}, 015006 (2008).

\bibitem{mew}
Moore, K.~W., Chakrabarti, R., Riviello, G., and Rabitz, H.
\newblock {\em Phys. Rev. A}{ \bf 83}, 012326 (2011).

\bibitem{greg}
Riviello, G., Chakrabarti, R., Moore, K.~W., and Rabitz, H.
\newblock in preparation,  (2011).

\end{thebibliography}
\newpage
{\bf Captions}

{\bf Table \ref{t1}.} Mean search effort, ratio $R_{\varepsilon}$, initial slope metric $\mathcal{S}_0$, and maximal slope metric $\mathcal{S}_{\max}$ for all simulations that showed invariant scaling effort to $N$. Recorded values are taken from simulations at $N$=20, but for other $N$ the values were similar. The values to the left of the double line are from simulations using the rotor $H_0$ (Eq. (\ref{ho})), and the values to the right of the double line are from simulations using the oscillator $H_0$ (Eq. (\ref{osc})). A comparison of the landscape metrics with the effort shows that the two are correlated. The ``easiest'' optimizations ($D$=1.0) have the lowest ratio $R_{\varepsilon}$ and the highest initial $\mathcal{S}_0$ and maximal $\mathcal{S}_{\max}$ slope metrics, while the ``hardest'' optimizations ($D$=0.2) have the highest ratios $R_{\varepsilon}$ and lowest initial and maximal slope metrics. The effort using the oscillator $H_0$ is always greater than for the rotor $H_0$, and the metrics show corresponding increases.

{\bf Figure 1.} Statistical distributions of $P_{i\to f}$ values for $N$=10, 15, and 20, with $D$=0.5 and $F$=10 for the $|1\rangle\to|N\rangle$ target. The inset depicts the mean value of distributions of initial $P_{i\to f}$ values for different dipoles, targets, and field parameters. The target transition, dipole drop off rate $D$ and field fluence $F$ are denoted as $P_{i\to f}$, $D$, $F$ in the legend. The mean initial value decreases for the $|1\rangle\to|N\rangle$ target, but is constant for fixed target transitions. Statistical error bars are shown for the $|1\rangle\to|N\rangle$ transition, and representative error bars for the other cases are shown as well. Some points are shifted on the x-axis for graphical clarity.

{\bf Figure 2.} Required mean search effort versus $N$ for the target transitions $|1\rangle\to|5\rangle$ (solid shapes) and $|1\rangle\to|10\rangle$ (open shapes) for Hamiltonians with dipole structures of $D$=1.0 (squares), $D$=0.5 (circles), $D$=0.2 (down triangles) and $\alpha$ (side triangles), with $H_0$ given by Eq. (\ref{ho}) (a) and by Eq. (\ref{osc}) (b) Search effort is invariant to $N$ in all cases (excepting some cases where the effort for the smallest $N$ recorded is significantly lower than for remaining $N$), but the absolute effort is greater for weak coupled dipoles, the $|1\rangle\to|10\rangle$ transition, and oscillator $H_0$ structure. Some points are shifted on the abscissa for graphical clarity.

{\bf Figure 3.} Population of states versus time for a 10-level system with target $|1\rangle\to|10\rangle$. (a) $D$=0.2, and all intermediate states $|2\rangle$ through $|9\rangle$ are accessed sequentially on the way from $|1\rangle$ to $|10\rangle$, consistent with a ladder-climbing mechanism. (b) $D$=1.0, and only states $|2\rangle$ and $|8\rangle$ play a significant role (all other intermediate states are never populated more than 10$\%$ and are not shown).

{\bf Figure 4.} Required mean search effort versus $N$ for the target transition $|1\rangle\to|5\rangle$ and $\alpha$ dipole structure with varying initial field strength and bandwidth. The strength (solid shapes) or bandwidth (open shapes) is labeled in the legend. For low strength and reasonable bandwidth, effort is invariant to $N$. For high fluence ($F$=100), effort scales exponentially with $N$, as shown by the least squares fit on the semi-log plot. For large bandwidth range $\Omega$, effort increases through $N$=20 and then levels off. Some points are shifted on the abscissa for graphical clarity.

{\bf Figure 5.} Plots of the absolute value of the matrix elements of the propagator $\{|U_{if}(T,0)|^2\}$ at initial random control fields (top) and optimal fields (bottom) for $N$=10 and the target $|1\rangle\to|5\rangle$ transition under the conditions $F$=1 (a) and $F$=100 (b). The (5,1) element is circled in each plot and has a value of 1.0 at the optimal fields and a value of close to zero at the initial fields. Both the initial and final $\{|U_{if}(T,0)|^2\}$ matrices are nearly diagonal for $F$=1, while at $F$=100, many off-diagonal elements are non-zero, indicating that the population is spread out among many states. Such a $\{|U_{if}(T,0)|^2\}$ matrix structure at $F$=100 results in a greater search effort because it becomes more difficult to gather all of the amplitude in a single final target state.

{\bf Figure 6.} Mean search effort versus $N$ for the $|1\rangle\to|N\rangle$ transition. When all transitions are allowed ($D$=1.0, squares), the effort is invariant to $N$. When the coupling strength decreases with distance between the states ($\alpha$, triangles, and $D$=0.5, circles), the effort scales exponentially with $N$, as shown by the least squares fit lines on the semi-log plot. Results are qualitatively the same for the rotor Hamiltonian (filled symbols) and oscillator Hamiltonian (open symbols).

\begin{figure}
\includegraphics[width=8.5cm]{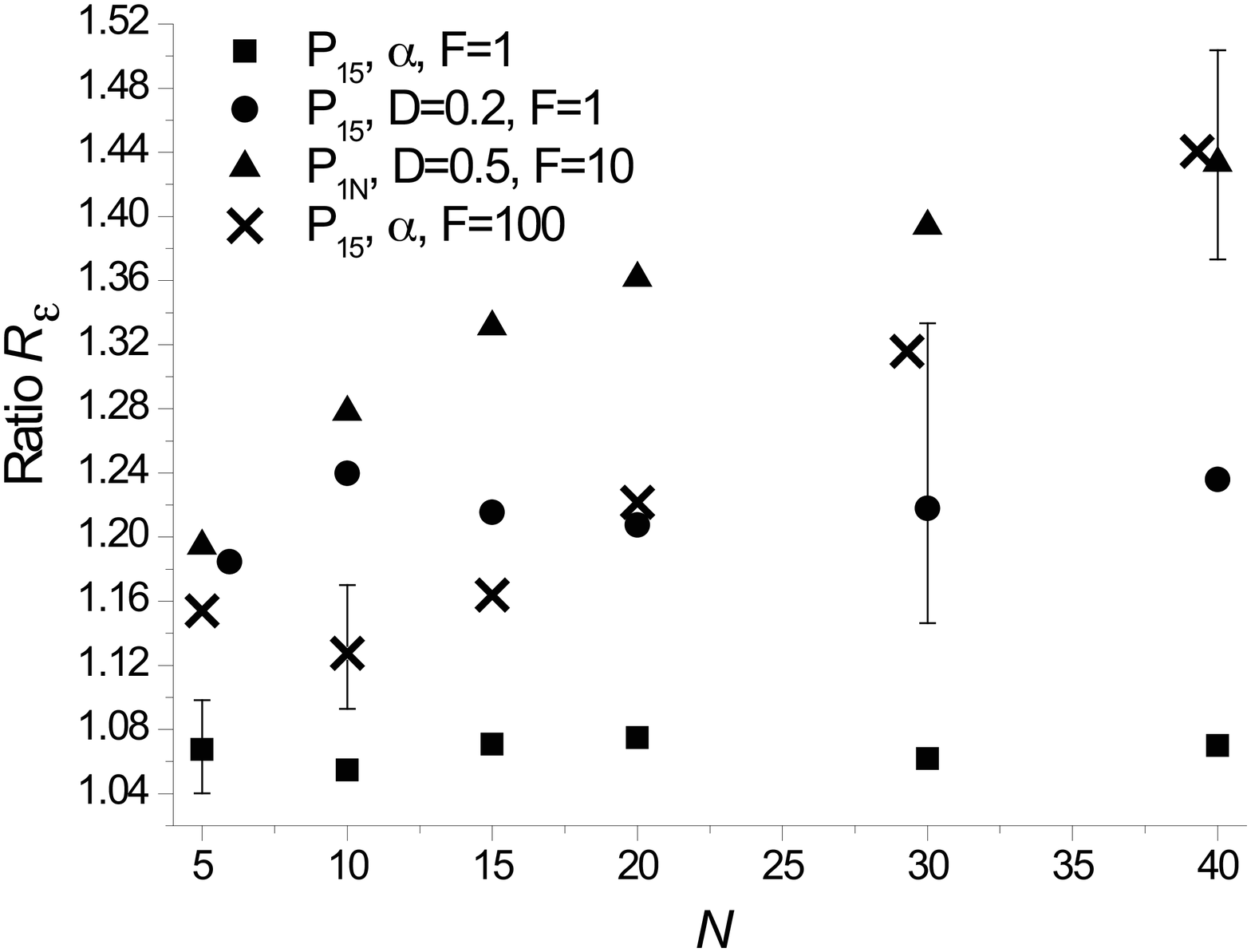}
\caption{\label{path}}
\end{figure}

{\bf Figure \ref{path}.} Ratio $R_{\varepsilon}$ of the control search path length to the Euclidian distance versus $N$ for selected optimizations from Section \ref{Scale}. The ratio is invariant to $N$ for the $|1\rangle\to|5\rangle$ transition and small strength $F$; the effort was also invariant to $N$ for these cases. $R_{\varepsilon}$ increases with $N$ for optimizations with $N$-dependent effort (i.e., $|1\rangle\to|N\rangle$ and $|1\rangle\to|5\rangle$ with $F$=100). Regardless of these variations, the values of $R_{\varepsilon}$ are generally close to 1, indicating that the searches follow direct trajectories in the space of controls.

\begin{figure}
\includegraphics[width=8.5cm]{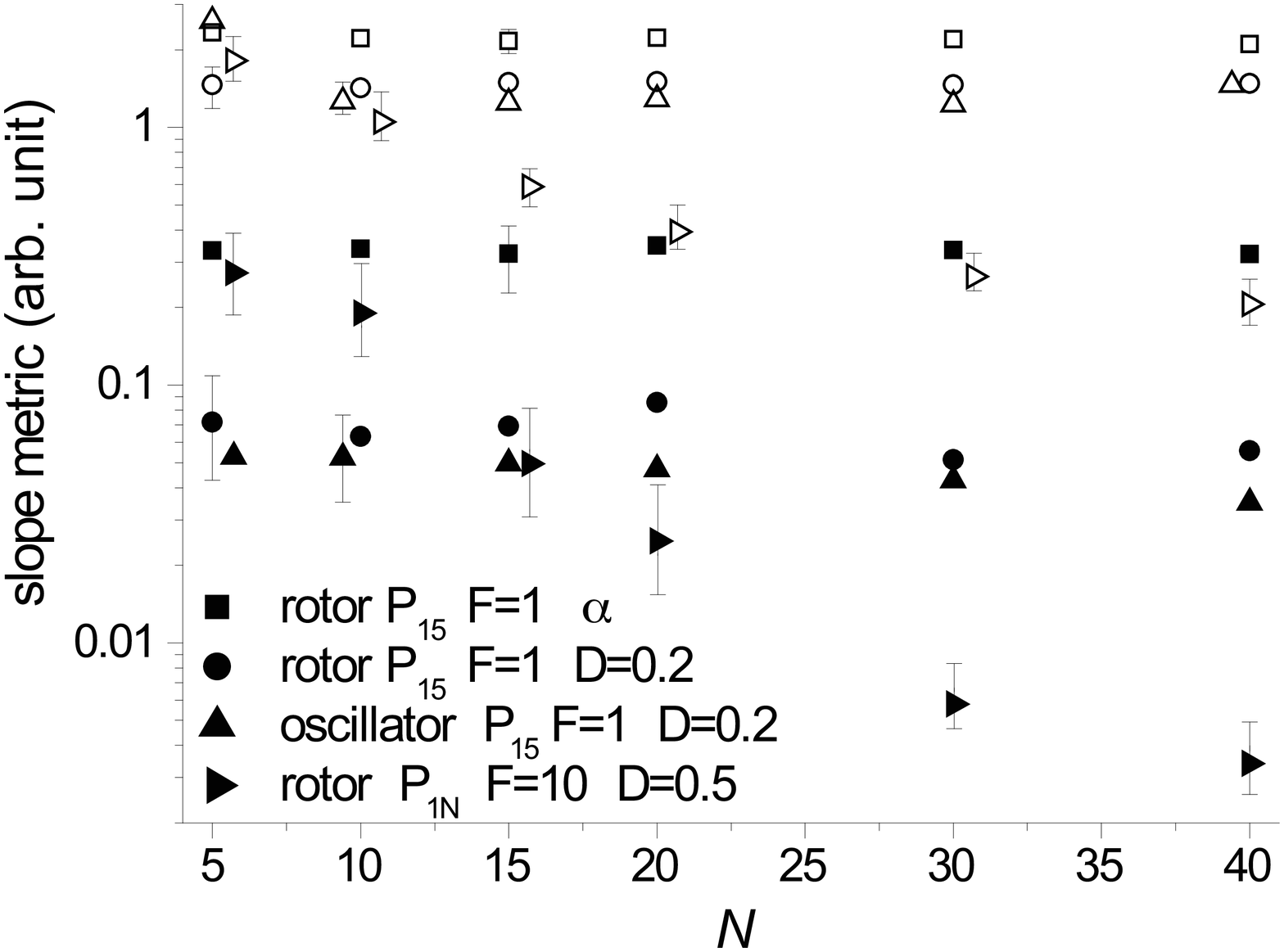}
\caption{\label{slope}}
\end{figure}

{\bf Figure \ref{slope}.} Initial (filled symbols) slope metric $\mathcal{S}_0$ and maximal (open symbols) slope metric $\mathcal{S}_{max}$ versus $N$ for selected cases from Section \ref{Scale}. With the exception of the $|1\rangle\to|N\rangle$ transition, the initial and final slope metrics are invariant to $N$, in agreement with the observed scaling behavior. Some points are shifted on the abscissa for graphical clarity.

\begin{figure}
 \includegraphics[width=8.5cm]{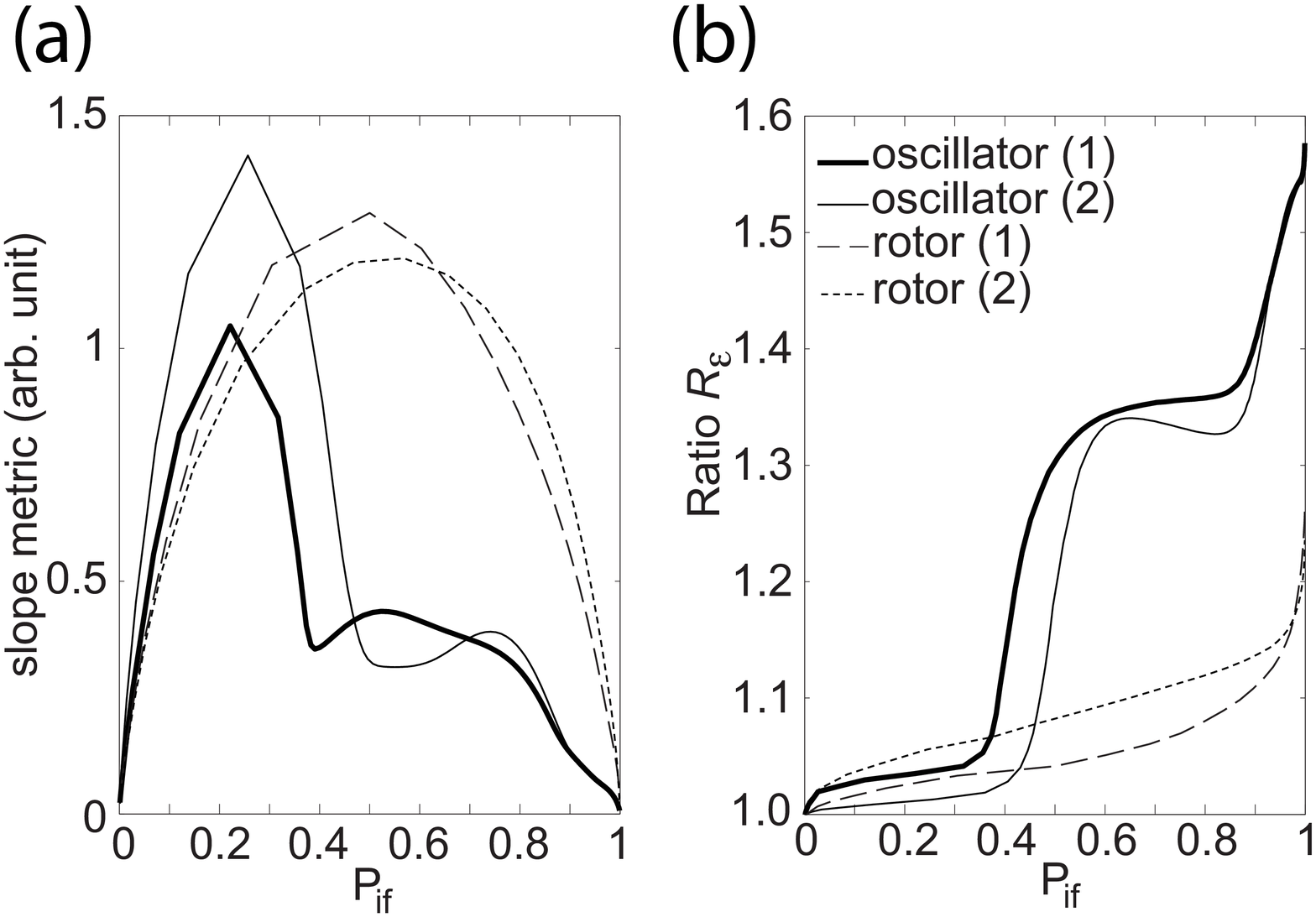}
\caption{\label{traject}}
\end{figure}

{\bf Figure \ref{traject}.} Trajectory of the slope metric (a) and ratio $R_\varepsilon$ (b) for two searches using the oscillator $H_0$ (solid lines) and rotor $H_0$ (dashed lines). $D$=0.2, $F$=1 and the target transition is $|1\rangle\to|5\rangle$. The trajectories for searches using the rotor $H_0$ are less complex than those for searches using the oscillator $H_0$.

\begin{figure}
\includegraphics[width=8.5cm]{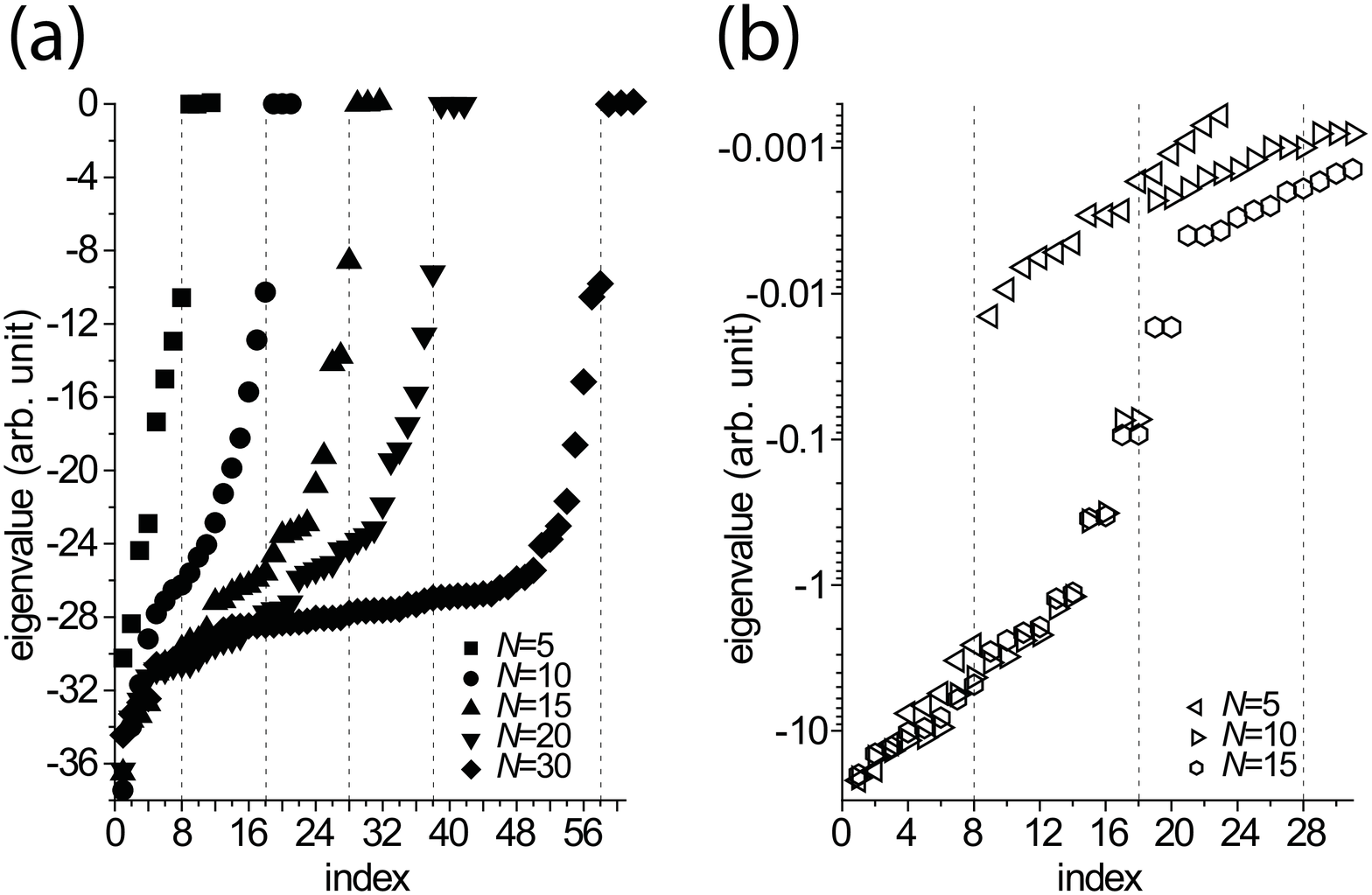}
\caption{\label{eigs}}
\end{figure}

{\bf Figure \ref{eigs}.} Hessian eigenvalues at the top of the landscape plotted versus their index. All optimizations used rotor $H_0$ structure and $F$=1 for the $|1\rangle\to|5\rangle$ transition. (a) optimizations with $D$=1.0. (b) optimizations with $D$=0.5; note the logarithmic scale. The vertical dotted lines show the value of 2$N$-2 for each $N$, and the 2$N$-2 rule is obeyed. In each case a few of the zero eigenvalues are shown for graphical clarity.
\renewcommand{\baselinestretch}{1}

\begin{table}[b]
\begin{tabular}{|c|c|c|c|c|c|c|c||c|c|c|c|}
\hline
$|i\rangle\to|f\rangle$&$\mu$&$F$&$\Omega$&Effort (\ref{ho})&$R_{\varepsilon}$&${\cal S}_0$&${\cal S}_{max}$&Effort (\ref{osc})&$R_{\varepsilon}$&${\cal S}_0$&${\cal S}_{max}$\\\hline
$|1\rangle\to|5\rangle$&$D$=1.0&1&$\omega_{5}$&17&1.07&0.49&3.17&21&1.10&0.27&2.99\\\hline
&$\alpha$&&&18&1.07&0.35&2.23&39&1.23&0.10&1.12\\\hline
&$D$=0.5&&&27&1.15&0.15&1.48&102&1.49&0.12&1.24\\\hline
&$D$=0.2&&&39&1.21&0.09&1.50&331&1.54&0.05&1.29\\\hline
$|1\rangle\to|10\rangle$&$D$=1.0&1&$\omega_{10}$&17&1.05&0.41&2.94&19&1.07&0.34&2.84\\\hline
&$\alpha$&&&21&1.07&0.13&1.44&52&1.13&0.05&0.59\\\hline
&$D$=0.5&&&38&1.21&0.02&0.83&113&1.37&0.01&0.71\\\hline
&$D$=0.2&&&118&1.38&0.02&0.82&891&1.69&0.01&0.60\\\hline
$|1\rangle\to|5\rangle$&$\alpha$&0.1&$\omega_{5}$&18&1.06&0.10&2.22&&&&\\\hline
&&10&&24&1.13&0.38&2.15&&&&\\\hline
&&1&$\omega_{20}$&26&1.1&0.11&1.92&&&&\\\hline
&&&$\frac{\omega_{N}}{2}$&27&1.11&0.12&1.97&&&&\\\hline
 \end{tabular}
\caption{\label{t1}}
\end{table}

\newpage
\newpage
\newpage
\newpage
\newpage
\newpage
\newpage
\newpage
\newpage
\newpage
\end{document}